\documentclass{article}

\PassOptionsToPackage{numbers, compress}{natbib}

 \usepackage[preprint]{neurips_2026}

\usepackage[utf8]{inputenc} 
\usepackage[T1]{fontenc}    
\usepackage{hyperref}       
\usepackage{url}            
\usepackage{booktabs}       
\usepackage{amsfonts}       
\usepackage{nicefrac}       
\usepackage{microtype}      
\usepackage[table]{xcolor}  
\usepackage{graphicx}       
\usepackage[most]{tcolorbox}  
\definecolor{model1}{RGB}{218, 218, 255}
\definecolor{model2}{RGB}{255, 235, 205}
\definecolor{rankfirst}{RGB}{189, 215, 238}
\definecolor{ranksecond}{RGB}{198, 239, 206}
\definecolor{mygreen}{RGB}{85,100,40}
\definecolor{softblue}{RGB}{90, 144, 180}
\definecolor{linkblue}{RGB}{41, 98, 155}
\hypersetup{
   colorlinks = true,
    linkcolor = linkblue,
     urlcolor = linkblue,
    citecolor = mygreen,
    filecolor = blue,
   }

\newtcolorbox{promptbox}[1][]{%
  colback=gray!5, colframe=black!60, boxrule=0.5pt, arc=2pt,
  left=5pt, right=5pt, top=4pt, bottom=4pt,
  fonttitle=\bfseries\sffamily\small,
  fontupper=\small\rmfamily,
  breakable, #1
}

\title{Trust or Abstain? A Self-Aware RAG Approach}

\author{%
  Xi Zhu$^{\,1}$
  \quad Ziqi Wang$^{\,1}$
  \quad Kai Mei$^{\,1}$
  \quad Wujiang Xu$^{\,1}$
  \quad Minghao Guo$^{\,1}$ \\[0.2em]
  \textbf{Bangji Yang}$^{\,2}$
  \quad \textbf{Jiajun Fan}$^{\,2}$
  \quad \textbf{Dimitris N. Metaxas}$^{\,1}$ \\[0.5em]
  $^{1}$Rutgers University \qquad $^{2}$University of Illinois Urbana-Champaign \\
}

\begin{document}

\begingroup
\renewcommand{\thefootnote}{}
\footnotetext{Email: xi.zhu@rutgers.edu}
\endgroup

\maketitle

\begin{abstract}
\label{abstract}
Retrieval-augmented generation (RAG) improves large language models (LLMs) by incorporating external evidence, but it also introduces knowledge conflicts when retrieved contextual knowledge (CK) and parametric knowledge (PK) disagree or are both unreliable. Existing approaches mainly coordinate which source to use, without explicitly asking whether each answer path is correct. We argue that faithful RAG requires LLM self-awareness, namely the ability to recognize the limits of its own knowledge and reasoning. To ground this problem, we construct a model-specific, ground-truth-aligned knowledge-conflict benchmark by evaluating LLM backbones on PK-only and CK-conditioned answer paths over approximately 69K query-context instances per backbone, drawn from five conflict-QA datasets. We then introduce SABER, a Self-Aware Belief Estimator for RAG that requires no LLM fine-tuning. SABER combines a self-prior with PK-side and CK-side conditional reasoning representations from multi-trace inference, then estimates reliability beliefs with two lightweight predictors to drive a 4-cell decision over trust PK, trust CK, trust either, or abstain. Across four LLM backbones, SABER improves end-to-end accuracy and conflict-specific faithfulness over ten inference-time and fine-tuning baselines, with the largest gains on conflict-heavy datasets. Under abstention, SABER's risk-coverage curve Pareto-dominates every prompt-based abstainer, providing a tunable balance between coverage and answer risk. Our code is available at \url{https://github.com/xizhu1022/SABER}.
\end{abstract}

\section{Introduction}
\label{main:intro}
Retrieval-augmented generation (RAG) \citep{lewis2020rag,wang2026ragrouter} is a widely-used inference-time strategy that supplies a large language model (LLM) with external evidence, complementing its parametric knowledge (PK) with retrieved contextual knowledge (CK). However, PK and CK are not always consistent, as they may agree, disagree, or both fail. This gives rise to a knowledge-conflict problem in which the LLM must decide how to use PK and retrieved CK when generating a faithful answer \citep{longpre2021entity,xie2024adaptive,wu2025clasheval,mallen2023nottrust,xu2024kcsurvey}.

\begin{figure}[t]
\centering
\vspace{-8mm}
\includegraphics[width=\linewidth]{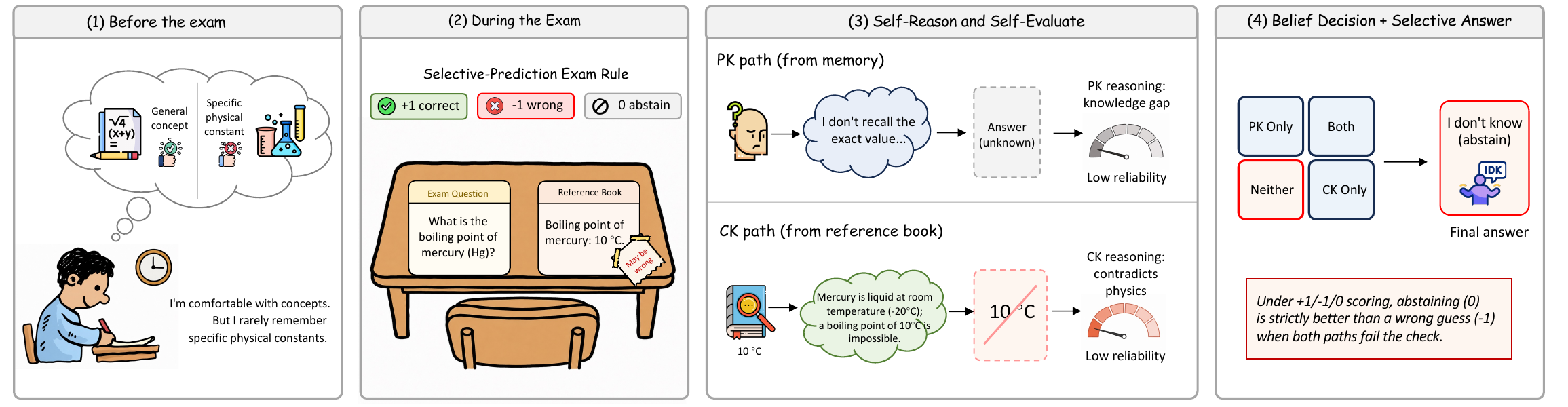}
\caption{A student facing a $+1$ / $-1$ / 0 exam rule, where a correct answer earns $+1$, a wrong one $-1$, and abstention earns $0$. The student’s memory corresponds to PK, while the potentially flawed reference book corresponds to retrieved CK. The example highlights two forms of self-awareness, namely knowledge-boundary awareness, which reveals when internal knowledge is insufficient, and reasoning-reliability awareness, which detects when retrieved context leads to an implausible answer. When both paths are unreliable, abstention is preferable to an unsupported prediction.}
\label{fig:pre_study}
\vspace{-4mm}
\end{figure}

Existing approaches follow three broad trajectories on how to choose between or fuse PK and CK. Source-arbitration methods decide which source should be trusted \citep{wu2025clasheval,shi2024cad,huang2025trust}. In parallel, context-quality methods determine whether the CK should be filtered or repaired \citep{yu2024tacs,zhang2025faithfulrag,gao2025clear,jin2025massive}. Retrieval-regulation methods consider whether and how CK should be invoked \citep{asai2024selfrag,jeong2024adaptiverag,su2024dragin,yao2025seakr,guo2026deepsieve}. Despite these efforts, they do not explicitly and independently ask whether the PK-based answer path and the CK-based answer path would each produce a correct answer with respect to the ground truth. This omission creates a conceptual blind spot that excludes the both-unreliable outcome from the decision space, pushing the model toward a confidently incorrect and unsupported response rather than an \textbf{honest abstention}, i.e., \emph{I don't know}, when neither answer path is reliable.

An open-book exam analogy in Figure~\ref{fig:pre_study} illustrates how humans make such decisions. Before the exam, a student has a rough sense of their own knowledge, including what they know and how certain they are. This corresponds to knowledge-boundary awareness. During the exam, the student may answer from memory or consult a reference book to form an answer. Since the reference book may be flawed, the student must judge which answer path, if any, is likely to lead to a correct answer. This corresponds to reasoning-reliability awareness. If only one path seems reliable, the student should follow that path. If both seem reliable, either can work. If neither does, the honest response is to admit uncertainty or ignorance rather than make an unsupported guess.

In the RAG setting, the LLM acts as the student, PK as memory, and CK as the reference book. Without such self-assessment, two failures recur. Confident hallucination, a knowledge-boundary failure, arises when the model overestimates PK, dismisses corrective evidence, and hallucinates from memory \citep{mallen2023nottrust,wu2025clasheval,chen2024inside}. Context over-reliance, a reasoning-reliability failure, arises when the model over-relies on CK and adopts misleading retrieved context even when its own knowledge and reasoning could expose the error \citep{xie2024adaptive,huang2025trust}. We trace both vulnerabilities to a self-awareness gap, where the model lacks a knowledge-boundary self-prior that reflects what it actually knows \citep{yin2023dontknow} and a reasoning-reliability self-evaluation that assesses whether its reasoning under a given source is sound. Faithful RAG therefore requires more than source coordination. It requires the model to recognize the boundaries and reliability of its own knowledge and reasoning before committing to a final answer.

To close this gap, we first construct a ground-truth-aligned knowledge-conflict benchmark by running four LLM backbones under PK-only and CK-conditioned answer paths, recording the observed correctness against the ground truth for approximately 69K query-context instances per backbone, drawn from five conflict-QA datasets. Then, we introduce \textbf{Self-Aware Belief Estimator for RAG (SABER)}. Mirroring the open-book exam mechanism above, SABER decomposes self-awareness into a knowledge-boundary self-prior and a reasoning-reliability self-evaluation. It first extracts the self-prior representation from the query-only hidden state to capture the model's inherent knowledge boundary. SABER then generates and self-evaluates multiple reasoning traces under PK and CK, aggregating them through multi-trace test-time inference into robust PK- and CK-conditional states. Two lightweight predictors combine the self-prior with the PK-side and CK-side states to estimate their reliability beliefs, which drive a 4-cell decision over PK, CK, both, or neither. SABER then commits to the more reliable PK or CK answer, or honestly abstains when neither path is reliable.

Our contributions are summarized as follows.

$\bullet$ \textbf{{Problem Formulation.}} We formulate knowledge conflict as a ground-truth-aligned decision problem. Instead of choosing between PK and CK, we independently evaluate the correctness of the PK-only and CK-conditioned paths, yielding a 4-cell decision space over PK, CK, both, and neither.

$\bullet$ \textbf{{Benchmark Construction.}} We construct a benchmark with approximately 69K query--context instances per backbone across four LLM backbones and five conflict-QA datasets. Each instance records observed PK-path and CK-path correctness against the ground truth. We further introduce three conflict-specific faithfulness metrics to support controlled evaluation of path-level reliability and reveal failure modes hidden by overall accuracy.

$\bullet$ \textbf{{Method.}} We introduce SABER, a lightweight self-aware belief reasoning framework requiring no LLM fine-tuning. SABER combines a self-prior with PK- and CK-side conditional reasoning representations from multi-trace inference, then estimates reliability beliefs to drive a 4-cell decision over trust PK, trust CK, trust both, or abstain.

$\bullet$ \textbf{{Empirical Findings.}} SABER improves end-to-end accuracy and conflict-specific faithfulness over ten inference-time and trained baselines, with the largest gains on conflict-heavy datasets. Under abstention, SABER's risk-coverage curve Pareto-dominates every prompt-based abstainer, providing a tunable threshold that lets practitioners trade coverage against answer risk without retraining.

\section{A Knowledge-Conflict Benchmark for LLM Self-Awareness}
\label{sec:benchmark}

LLM self-awareness cannot be studied without observing actual model behavior. Therefore, we construct a model-specific benchmark in which each instance is labeled by observed answer correctness. For every backbone, we record answers produced under a closed-book PK path and an open-book CK path, then label each path by whether it matches the ground truth.

\textbf{Datasets and LLM Backbones.} We construct the query-context pool from five conflict-QA datasets, including ConFiQA \citep{bi2024confiqa} with three sub-tasks (QA, MR, MC), ConflictQA-PopQA \citep{xie2024adaptive} (abbr.\ ConflictQA), ConflictBank \citep{su2024conflictbank}, and the TriviaQA and Natural Questions (NQ) subsets prepared by \citet{huang2025trust}. We evaluate four LLM backbones on this pool, namely Qwen-2.5-3B-Instruct, Qwen-2.5-7B-Instruct, Llama-3.1-8B-Instruct, and Llama-3.2-3B-Instruct, yielding approximately 69K ground-truth-aligned PK/CK instances per backbone. Processing details, per-dataset split sizes, the 4-cell distribution, and the labeling prompts are in Appendices~\ref{sec:appendix-datasets} and~\ref{sec:appendix-prompt-construction}.

\textbf{Grounding PK and CK in Answer Correctness.} For each LLM backbone \(\mathcal{M}\) and instance \((q, c, a^\star)\), consisting of a query, a retrieved context, and a ground-truth answer or alias set, we generate two candidate answers. The PK-only answer is produced from the query alone, while the CK-conditioned answer is produced from both the query and the retrieved context.
\begin{equation}
a_{\mathrm{PK}} = \mathcal{M}(\mathrm{prompt}_{\mathrm{PK}}(q)), \qquad
a_{\mathrm{CK}} = \mathcal{M}(\mathrm{prompt}_{\mathrm{CK}}(q,c)).
\label{eq:answer-paths}
\end{equation}
We then evaluate both answers against the ground truth using an alias-aware matching function.
\begin{equation}
y_{\mathrm{PK}} = \mathrm{match}(a_{\mathrm{PK}}, a^\star), \qquad
y_{\mathrm{CK}} = \mathrm{match}(a_{\mathrm{CK}}, a^\star).
\label{eq:correctness-labels}
\end{equation}
The pair \((y_{\mathrm{PK}}, y_{\mathrm{CK}})\) partitions the benchmark into four reliability cells \(C_{ab}\), where \(a,b \in \{0,1\}\) indicate the correctness of the PK and CK paths. These cells correspond to both paths correct \((C_{11})\), only PK correct \((C_{10})\), only CK correct \((C_{01})\), and neither path correct \((C_{00})\). These cells provide a ground-truth basis for studying whether the model can form accurate beliefs about its self-awareness.

\textbf{Evaluation Metrics.} Beyond \textbf{end-to-end answer accuracy}, we use three conflict-specific faithfulness metrics. \textbf{CF (Context Faithfulness)} measures accuracy on instances where only the CK-conditioned answer is correct, testing whether a method follows reliable CK when PK is misleading. \textbf{KF (Knowledge Faithfulness)} measures accuracy on instances where only the PK-only answer is correct, testing whether a method preserves correct PK when CK is misleading. \textbf{MFS (Macro Faithfulness Score)} averages CF and KF, rewarding methods that handle both conflict directions. The selective-prediction setting in \S\ref{sec:abstention} additionally evaluates calibrated refusals using Score, Coverage, Risk on answered (\(R_C\)), and abstention F1 (\(F_1\)). Formal definitions appear in Appendix~\ref{sec:appendix-metrics}.

\section{Method: SABER}
\label{sec:method}

\begin{figure}[t]
\centering
\vspace{-5mm}
\includegraphics[width=\linewidth]{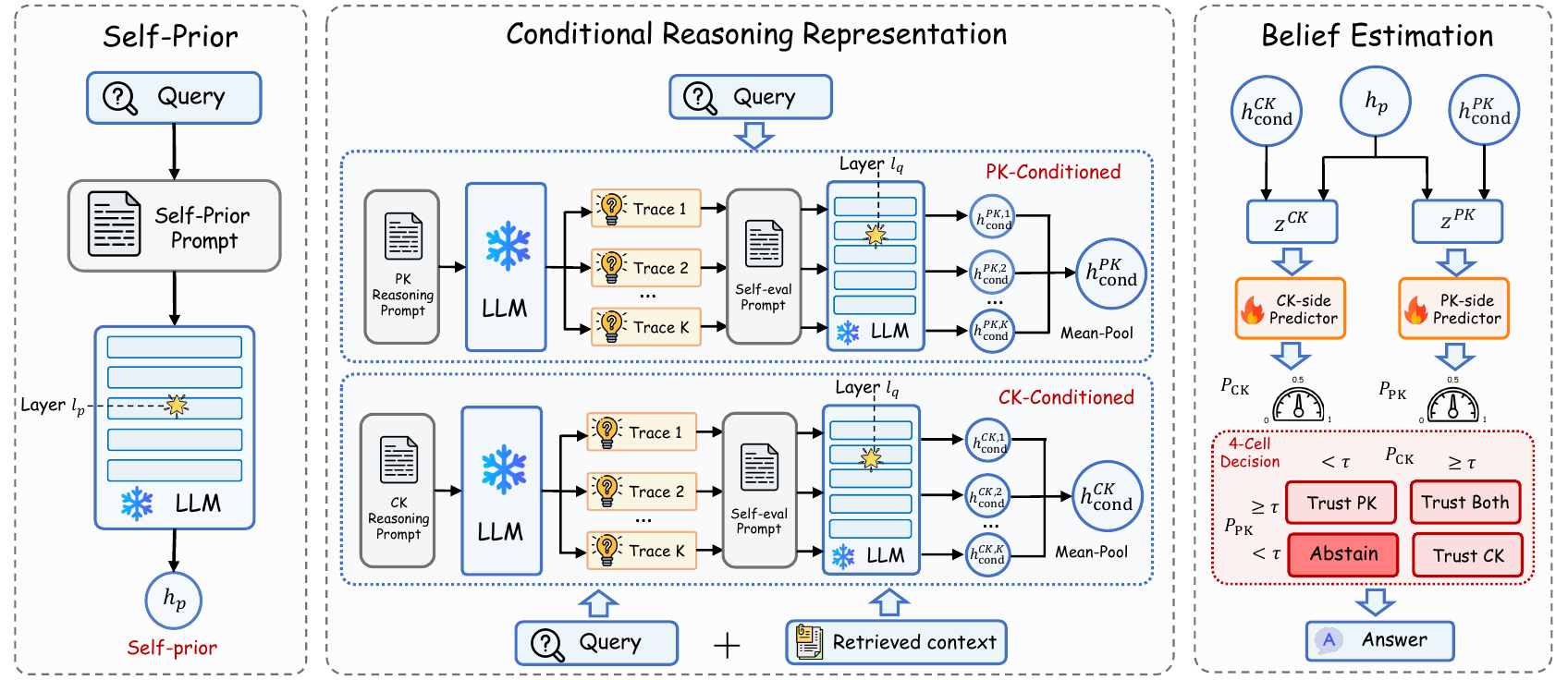}
\caption{SABER pipeline. The frozen LLM produces a query-only self-prior and per-side conditional reasoning representations via multi-trace generation and self-evaluation. Two lightweight predictors output CK- and PK-beliefs that drive the 4-cell decision with optional abstention.}
\label{fig:main_figure}
\vspace{-3mm}
\end{figure}

SABER is a lightweight two-stage belief estimator that operates at test time without fine-tuning the backbone LLM. Building on the benchmark, it extracts a self-prior representation from the frozen LLM (\S\ref{sec:method-prior}) and per-side conditional reasoning representations (\S\ref{sec:method-cond}), then feeds them into two lightweight predictors that estimate reliability beliefs over PK-side and CK-side correctness. These beliefs drive an internal 4-cell decision that produces SABER's final response, including explicit abstention when neither side is reliable (\S\ref{sec:method-decision}). The full pipeline is illustrated in Figure~\ref{fig:main_figure}.

\subsection{Self-Prior Representation}
\label{sec:method-prior}

The self-prior representation captures the LLM backbone's internal confidence before any retrieved context or reasoning trace is introduced. It reflects the model's self-assessed competence on the current query, signaling whether later reasoning should rely on PK or seek support from CK. Following prior work that reads implicit beliefs from frozen-LLM hidden states \citep{kadavath2022language, burns2022discovering, azaria2023internal}, we extract this signal by feeding the query alone into the frozen LLM $\mathcal{M}$ and reading the hidden state at a fixed prior layer $\ell_p$,
\begin{equation}
h_q = \phi^{(\ell_p)}_{\mathcal{M}}\!\bigl(\mathrm{prompt}_{\mathrm{prior}}(q)\bigr).
\label{eq:self-prior}
\end{equation}
The self-prior prompt is in Appendix~\ref{sec:appendix-prompt-prior}). This mirrors a student's pre-exam sense of whether memory can answer the question or a potentially flawed reference book is needed, shaping whether the book is treated as auxiliary evidence or relied on more heavily. The self-prior therefore conditions both PK and CK predictions without determining either one.

\subsection{Conditional Reasoning Representation}
\label{sec:method-cond}

The conditional reasoning representation captures how the model reasons under each source assumption and how it self-evaluates that reasoning after generation. Unlike the self-prior, which reflects pre-reasoning confidence, this representation provides a post-reasoning self-awareness signal. We obtain this signal through a generate-then-evaluate procedure, where the model first generates a source-conditioned reasoning trace and then runs a self-evaluation forward pass over that trace.

\textbf{Source-conditional Reasoning Trace.} For each source, we let the same LLM sample chain-of-thought reasoning traces and answers under the corresponding knowledge condition. On the PK side, the model answers in a closed-book setting from its own knowledge. On the CK side, the model anchors its reasoning only on the retrieved context. We denote the $k$-th sampled trace under source $s \in \{\mathrm{PK}, \mathrm{CK}\}$ by $t_{s,k}$; the resulting trace provides a side-specific view of how the model walks toward an answer, revealing which knowledge it draws on and how it integrates or resists the available evidence. The exact reasoning prompts are given in Appendix~\ref{sec:appendix-prompt-pk} and Appendix~\ref{sec:appendix-prompt-ck}.

\textbf{Self-evaluation Forward Pass.} To estimate confidence in the generated reasoning trace, we ask the LLM to self-evaluate it in a second forward pass, indicating whether the trace is likely to support a correct answer under the assumed source \citep{kadavath2022language, mei2025rwom}. Specifically, we read per-token hidden states at a fixed conditional layer \(\ell_c\) over the reasoning span \citep{ni2026reprobe} and mean-pool them into a per-trace vector,
\begin{equation}
h^{(s,k)}_{\mathrm{cond}} = \bar{\phi}^{(\ell_c)}_{\mathcal{M}}\!\bigl(\mathrm{prompt}_{\mathrm{self\text{-}eval}}(q, c, t_{s,k})\bigr),
\label{eq:cond-pertrace}
\end{equation}
where \(t_{s,k}\) denotes the \(k\)-th reasoning trace under source \(s\), and \(\bar{\phi}^{(\ell_c)}_{\mathcal{M}}(\cdot)\) averages hidden states at layer \(\ell_c\) over the reasoning span. This self-evaluation may reflect the trace's coherence, internal consistency, and use of available knowledge, which together shape the LLM's confidence in the trace. It mirrors how students, when uncertain about memory or a flawed reference book, reason through a problem and judge whether the answer is plausible. The self-evaluation prompt is in Appendix~\ref{sec:appendix-prompt-eval}.

\textbf{Multi-trace Test-time Inference.} A single trace is one sample from a stochastic generator and may run into a dead end, hallucinate a step, or react to surface features of $c$. Following multi-trace test-time inference \citep{wang2023selfconsistency}, we sample $K$ independent reasoning traces per side and mean-pool their per-trace vectors into a robust per-side conditional state,
\begin{equation}
h^{(s)}_{\mathrm{cond}} = \mathrm{MeanPool}\bigl(h^{(s,1)}_{\mathrm{cond}},\, h^{(s,2)}_{\mathrm{cond}},\, \ldots,\, h^{(s,K)}_{\mathrm{cond}}\bigr), \quad s \in \{\mathrm{PK}, \mathrm{CK}\}.
\label{eq:cond-perside}
\end{equation}
\subsection{Belief Estimation and Decision with Abstention}
\label{sec:method-decision}

This section turns SABER's two self-awareness signals into an actionable belief decision. We first combine the self-prior and conditional reasoning representations into joint representations for the PK and CK sides. We then estimate separate PK-side and CK-side reliability beliefs using two lightweight predictors and map them into a 4-cell decision space with explicit abstention.

\textbf{Joint Representation.} For each side $s$, we combine the self-prior with the corresponding conditional reasoning representation to form a side-specific joint representation,
\begin{equation}
z^{(s)} = f_{\mathrm{fuse}}\!\bigl(h_q,\, h^{(s)}_{\mathrm{cond}}\bigr), \quad s \in \{\mathrm{PK}, \mathrm{CK}\},
\label{eq:posterior}
\end{equation}
where $f_{\mathrm{fuse}}$ concatenates the self-prior with the side-specific reasoning representation. Notably, the same self-prior plays complementary roles across the two sides. For PK, it reflects the model's initial confidence in its own parametric knowledge. For CK, it reflects how much the model may need to lean on external evidence. Combined with the corresponding source-conditioned reasoning state, this prior yields a side-specific joint representation used to estimate the reliability of each answer path.

\textbf{Per-side Correctness Predictor.} Because SABER's self-awareness signals are anchored in ground-truth answer correctness, the PK-side and CK-side paths give rise to two independent correctness questions whose outcomes span four possible combinations. Along this line, we employ two lightweight correctness predictors, one for each side, each taking the concatenation of the two joint representations as input and outputting a scalar reliability belief,
\begin{equation}
\hat p_s = f_{\mathrm{pred}, s}\!\bigl([z^{(\mathrm{PK})};\, z^{(\mathrm{CK})}]\bigr), \quad s \in \{\mathrm{PK}, \mathrm{CK}\}.
\label{eq:reliability}
\end{equation}
The shared input allows each predictor to draw on evidence from both sides, such as using CK-side unreliability to sharpen the PK-side judgment, while the independent outputs preserve all four reliability outcomes and feed directly into the 4-cell decision space. The two predictors are trained independently with binary cross-entropy on \(y_{\mathrm{PK}}\) and \(y_{\mathrm{CK}}\) from our constructed ground-truth-aligned benchmark. The LLM backbone remains frozen throughout. Only the parameters of predictors are updated, adding few trainable parameters and negligible inference cost. Architecture and training hyperparameters of the two predictors, together with the multi-trace generation parameters used to produce the conditional reasoning traces, are reported in Appendix~\ref{sec:appendix-implementation}.

\textbf{4-Cell Decision with Abstention.} Thresholding the two reliability beliefs \((\hat p_{\mathrm{PK}}, \hat p_{\mathrm{CK}})\) at \(\tau\) assigns each instance to one of the four cells defined in \S\ref{sec:benchmark}. For the three reliable cells, SABER returns the PK answer on \(C_{10}\), the CK answer on \(C_{01}\), or the candidate with the higher reliability belief on \(C_{11}\), all without invoking the LLM again. The \(C_{00}\) cell, where neither side is reliable, supports two deployment modes. The default mode falls back to the higher-belief candidate, preserving full coverage for low-stakes settings, while the abstention mode returns \emph{I don't know}, suiting high-stakes settings where unsupported answers are costly. The threshold \(\tau\) controls how often SABER commits to an answer and, under abstention, trades coverage against answer risk. As shown in \S\ref{sec:abstention}, sweeping \(\tau\) produces a continuous risk-coverage curve that existing methods cannot match, making \(\tau\) a deployment-relevant control knob rather than a purely internal hyperparameter.

\section{Experiments}
\label{sec:experiments}


We run experiments on the five datasets in our ground-truth-aligned benchmark from \S\ref{sec:benchmark}.
We compare against ten methods from four categories. \textbf{(1) Vanilla baselines.} Vanilla LLM (Closed-Book) and Vanilla RAG. \textbf{(2) Verbal self-evaluation.} Internal-Eval, Context-Eval, Implicit-SCR, and Explicit-SCR \citep{huang2025trust}, which prompt the LLM to judge source reliability or its own confidence. \textbf{(3) Probability- and probe-based heuristics.} TPC \citep{wu2025clasheval} which compares the token-level probability, and TACS-LR \citep{yu2024tacs} that uses a linear probe for context filtering. \textbf{(4) Trained methods.} CR-DPO \citep{huang2025trust}, a DPO-style extension of Explicit-SCR, and R-Tuning \citep{zhang2024rtuning}, which fine-tunes the LLM with uncertain instances. Descriptions are provided in Appendix~\ref{sec:appendix-baselines}.
We evaluate all methods using the benchmark metrics from \S\ref{sec:benchmark}, namely end-to-end answer accuracy (Acc), Context Faithfulness (CF), Knowledge Faithfulness (KF), and Macro Faithfulness Score (MFS). Selective-answering metrics (Score, Coverage, R$_C$, abstention F$_1$) are introduced in \S\ref{sec:abstention} and used only there to study abstention decisions. Computational cost is reported in Appendix~\ref{sec:appendix-compute}.

\begin{table*}[t]
\centering
\vspace{-6mm}
\scriptsize
\setlength{\tabcolsep}{2pt}
\caption{Main results on Llama-3.1-8B and Llama-3.2-3B. We report Acc ($\uparrow$), CF, KF, and MFS ($\uparrow$), all as percentages, on five datasets. ConFiQA combines its three sub-tasks (QA / MR / MC) into a single dataset due to the limited space (breakdown is in Appendix~\ref{sec:appendix-main-full}). Since CF or KF alone can be saturated by always trusting one source, MFS is the primary conflict-faithfulness metric. For Acc and MFS, the best is highlighted with a {\setlength{\fboxsep}{1pt}\colorbox{rankfirst}{\textbf{blue}}} background and the second-best with a {\setlength{\fboxsep}{1pt}\colorbox{ranksecond}{green}} background.}
\label{tab:main-llama}
\resizebox{\textwidth}{!}{%
\begin{tabular}{l|cccc|cccc|cccc|cccc|cccc}
\toprule
 & \multicolumn{4}{c|}{\cellcolor{gray!15}\textbf{ConFiQA}} & \multicolumn{4}{c|}{\cellcolor{gray!15}\textbf{ConflictQA}} & \multicolumn{4}{c|}{\cellcolor{gray!15}\textbf{ConflictBank}} & \multicolumn{4}{c|}{\cellcolor{gray!15}\textbf{TriviaQA}} & \multicolumn{4}{c}{\cellcolor{gray!15}\textbf{NQ}} \\
Method & Acc$\uparrow$ & CF & KF & MFS$\uparrow$ & Acc$\uparrow$ & CF & KF & MFS$\uparrow$ & Acc$\uparrow$ & CF & KF & MFS$\uparrow$ & Acc$\uparrow$ & CF & KF & MFS$\uparrow$ & Acc$\uparrow$ & CF & KF & MFS$\uparrow$ \\
\midrule
\rowcolor{model1}
\multicolumn{21}{c}{\textbf{Llama-3.1-8B}} \\
\midrule
Closed-Book   & 31.2 & 7.9 & 100.0 & 53.9 & 22.7 & 3.0 & 100.0 & 51.5 & 23.5 & 0.6 & 100.0 & 50.3 & 80.8 & 16.7 & 100.0 & 58.3 & 41.3 & 6.3 & 100.0 & 53.2 \\
Vanilla RAG   & \cellcolor{ranksecond}{57.0} & 100.0 & 17.5 & 58.7 & 46.8 & 100.0 & 7.5 & 53.7 & 69.4 & 100.0 & 1.7 & 50.9 & 57.1 & 100.0 & 4.2 & 52.1 & 46.7 & 100.0 & 4.1 & 52.0 \\
\midrule
Internal-Eval & 45.4 & 55.3 & 62.8 & 59.0 & 31.9 & 37.6 & 72.1 & 54.9 & 61.2 & 78.1 & 38.1 & 58.1 & 78.8 & 50.0 & 85.9 & 68.0 & 49.6 & 66.7 & 59.2 & 62.9 \\
Context-Eval  & 39.2 & 34.5 & 85.7 & 60.1 & 31.7 & 36.6 & 79.6 & 58.1 & 58.9 & 61.6 & 90.3 & \cellcolor{ranksecond}{76.0} & 75.4 & 62.5 & 73.2 & 67.9 & 45.9 & 52.4 & 67.3 & 59.9 \\
Implicit-SCR  & 53.7 & 78.6 & 43.9 & 61.3 & 47.3 & 89.5 & 36.1 & 62.8 & 29.3 & 39.6 & 5.1 & 22.4 & 65.0 & 91.7 & 36.6 & 64.1 & 50.4 & 88.9 & 32.7 & 60.8 \\
Explicit-SCR  & 40.0 & 45.6 & 75.3 & 60.5 & 40.9 & 73.9 & 52.4 & 63.1 & 22.2 & 23.8 & 28.7 & 26.3 & 63.3 & 83.3 & 56.3 & 69.8 & 46.3 & 73.0 & 53.1 & 63.0 \\
\midrule
TPC           & 55.1 & 86.6 & 60.5 & \cellcolor{ranksecond}{73.6} & 47.0 & 92.7 & 34.7 & 63.7 & \cellcolor{ranksecond}{69.6} & 92.3 & 36.4 & 64.3 & 68.8 & 91.7 & 40.8 & 66.3 & 51.7 & 93.7 & 36.7 & 65.2 \\
TACS-LR       & 43.6 & 59.7 & 30.5 & 45.1 & 42.9 & 90.1 & 12.2 & 51.2 & 33.8 & 46.8 & 3.7 & 25.2 & 53.8 & 87.5 & 5.6 & 46.6 & 47.5 & 92.1 & 20.4 & 56.2 \\
\midrule
R-Tuning      & 20.7 & 9.1 & 32.7 & 20.9 & 10.5 & 2.6 & 28.3 & 15.5 & 2.3 & 0.0 & 8.4 & 4.2 & 72.5 & 28.6 & 82.1 & 55.4 & 33.9 & 18.8 & 58.8 & 38.8 \\
CR-DPO        & 47.9 & 60.1 & 77.1 & 68.6 & \cellcolor{rankfirst}\textbf{53.8} & 94.9 & 81.6 & \cellcolor{ranksecond}{88.2} & 52.3 & 50.3 & 85.8 & 68.0 & \cellcolor{ranksecond}{84.2} & 95.8 & 88.7 & \cellcolor{rankfirst}\textbf{92.3} & \cellcolor{rankfirst}\textbf{61.2} & 79.4 & 73.5 & \cellcolor{rankfirst}\textbf{76.4} \\
\midrule
\textbf{SABER (Ours)} & \cellcolor{rankfirst}\textbf{58.4} & 92.8 & 72.2 & \cellcolor{rankfirst}\textbf{82.5} & \cellcolor{ranksecond}{53.5} & 97.0 & 85.0 & \cellcolor{rankfirst}\textbf{91.0} & \cellcolor{rankfirst}\textbf{80.2} & 97.0 & 88.1 & \cellcolor{rankfirst}\textbf{92.5} & \cellcolor{rankfirst}\textbf{85.2} & 66.7 & 98.6 & \cellcolor{ranksecond}{82.6} & \cellcolor{ranksecond}{52.1} & 73.0 & 65.3 & \cellcolor{ranksecond}{69.2} \\
\midrule
\rowcolor{model2}
\multicolumn{21}{c}{\textbf{Llama-3.2-3B}} \\
\midrule
Closed-Book   & 23.9 & 5.5 & 100.0 & 52.7 & 13.9 & 0.8 & 100.0 & 50.4 & 16.8 & 0.3 & 100.0 & 50.1 & \cellcolor{ranksecond}{73.3} & 3.2 & 100.0 & 51.6 & 36.4 & 7.6 & 100.0 & 53.8 \\
Vanilla RAG   & 50.3 & 100.0 & 9.7 & 54.8 & \cellcolor{ranksecond}{42.0} & 100.0 & 5.4 & 52.7 & 63.5 & 100.0 & 0.8 & 50.4 & 52.1 & 100.0 & 7.1 & 53.6 & 43.8 & 100.0 & 4.7 & 52.3 \\
\midrule
Internal-Eval & 41.1 & 62.3 & 60.7 & 61.5 & 37.2 & 74.1 & 65.0 & 69.5 & 59.6 & 81.0 & 62.1 & 71.5 & 72.1 & 61.3 & 78.6 & 69.9 & 43.8 & 56.1 & 67.4 & 61.8 \\
Context-Eval  & 46.6 & 77.4 & 63.7 & 70.6 & 29.5 & 51.9 & 68.1 & 60.0 & 49.9 & 61.5 & 82.3 & \cellcolor{ranksecond}{71.9} & 57.6 & 80.6 & 26.7 & 53.7 & 39.9 & 57.6 & 44.2 & 50.9 \\
Implicit-SCR  & 41.7 & 68.3 & 24.9 & 46.6 & 41.3 & 88.1 & 20.3 & 54.2 & \cellcolor{ranksecond}{64.7} & 93.3 & 7.3 & 50.3 & 48.2 & 83.9 & 11.5 & 47.7 & 43.0 & 91.0 & 11.4 & 51.2 \\
Explicit-SCR  & 28.1 & 32.8 & 70.4 & 51.6 & 27.6 & 54.6 & 68.5 & 61.6 & 27.8 & 33.2 & 50.0 & 41.6 & 52.1 & 61.3 & 42.9 & 52.1 & 30.6 & 60.6 & 27.9 & 44.3 \\
\midrule
TPC           & \cellcolor{ranksecond}{50.7} & 90.6 & 56.9 & \cellcolor{ranksecond}{73.7} & 41.4 & 92.4 & 37.8 & 65.1 & 57.0 & 85.0 & 24.2 & 54.6 & 67.9 & 100.0 & 48.8 & \cellcolor{ranksecond}{74.4} & \cellcolor{ranksecond}{47.5} & 93.9 & 30.2 & \cellcolor{ranksecond}{62.1} \\
TACS-LR       & 38.6 & 63.7 & 20.4 & 42.0 & 36.7 & 82.9 & 6.2 & 44.6 & 60.7 & 90.0 & 4.0 & 47.0 & 51.4 & 90.3 & 21.4 & 55.9 & 40.9 & 84.8 & 13.6 & 49.2 \\
\midrule
R-Tuning      & 3.0 & 0.9 & 7.0 & 4.0 & 0.4 & 0.0 & 0.9 & 0.5 & 0.2 & 0.0 & 0.8 & 0.4 & 42.5 & 15.1 & 55.9 & 35.5 & 23.1 & 10.9 & 45.7 & 28.3 \\
CR-DPO        & 34.6 & 41.7 & 81.5 & 61.6 & 39.8 & 79.2 & 85.6 & \cellcolor{ranksecond}{82.4} & 39.2 & 44.5 & 69.4 & 56.9 & 63.7 & 67.7 & 76.2 & 72.0 & 40.1 & 56.1 & 55.8 & 55.9 \\
\midrule
\textbf{SABER (Ours)} & \cellcolor{rankfirst}\textbf{54.1} & 94.8 & 79.4 & \cellcolor{rankfirst}\textbf{87.1} & \cellcolor{rankfirst}\textbf{46.9} & 97.9 & 82.9 & \cellcolor{rankfirst}\textbf{90.4} & \cellcolor{rankfirst}\textbf{70.9} & 97.6 & 82.3 & \cellcolor{rankfirst}\textbf{89.9} & \cellcolor{rankfirst}\textbf{77.9} & 77.4 & 85.7 & \cellcolor{rankfirst}\textbf{81.6} & \cellcolor{rankfirst}\textbf{54.1} & 90.9 & 69.8 & \cellcolor{rankfirst}\textbf{80.3} \\
\bottomrule
\end{tabular}}
\vspace{-2mm}
\end{table*}

\subsection{Main Results}
\label{sec:main-results}

Tables~\ref{tab:main-llama} and~\ref{tab:main-qwen} report Acc, CF, KF, and MFS on five datasets across four LLM backbones from Llama and Qwen. We highlight four observations.

\textbf{(1) SABER consistently outperforms all inference-time baselines on Acc and MFS.} Across the 20 backbone-dataset settings, SABER achieves the strongest Acc in all but one setting and the strongest MFS in every setting. Its gains over the best inference-time baseline reach $+$10.6\% Acc on (Llama-3.1-8B, ConflictBank) and $+$27.3\% MFS on (Llama-3.1-8B, ConflictQA). On average, SABER leads by $+$3.4\% in Acc and $+$15.4\% in MFS.

\textbf{(2) SABER is competitive with fully fine-tuned CR-DPO while keeping the LLM frozen.} CR-DPO's gains concentrate on larger LLMs, whereas SABER matches or exceeds CR-DPO in most settings and strictly dominates it on both 3B models across all datasets and metrics. This shows that SABER's hidden-state reliability signal can rival full DPO fine-tuning, especially on smaller models.

\textbf{(3) SABER's MFS gains substantially exceed its Acc gains, indicating that its main advantage is conflict resolution rather than raw accuracy.} Several baselines can obtain reasonable Acc by always favoring CK, as in Vanilla-RAG, or PK, as in Closed-Book, but they sacrifice either KF or CF. SABER maintains high CF and KF simultaneously, deciding when to trust context and when to preserve internal knowledge. Its only inference-time Acc loss on (Qwen-2.5-7B, ConFiQA) is offset by a $+$15.7\% MFS gain, illustrating the bias-vs-symmetry trade-off of prompt-only methods.

\textbf{(4) Acc gains are larger on Llama than on Qwen, while MFS gains remain uniform across families.} SABER's mean Acc lead is $+$4.9\% on Llama and $+$2.0\% on Qwen, while MFS gains stay between $+$14\% and $+$16\% on both families. The narrower Qwen Acc margin reflects the stronger Implicit-SCR baseline on Qwen rather than model scale, and the uniform MFS gain shows that SABER's conflict-resolution advantage transfers across families even when the Acc gap narrows.

\begin{table*}[t]
\centering
\vspace{-7mm}
\scriptsize
\setlength{\tabcolsep}{2pt}
\caption{Main results on Qwen-2.5-7B and Qwen-2.5-3B. Same setup as Table~\ref{tab:main-llama}.}
\label{tab:main-qwen}
\resizebox{\textwidth}{!}{%
\begin{tabular}{l|cccc|cccc|cccc|cccc|cccc}
\toprule
 & \multicolumn{4}{c|}{\cellcolor{gray!15}\textbf{ConFiQA}} & \multicolumn{4}{c|}{\cellcolor{gray!15}\textbf{ConflictQA}} & \multicolumn{4}{c|}{\cellcolor{gray!15}\textbf{ConflictBank}} & \multicolumn{4}{c|}{\cellcolor{gray!15}\textbf{TriviaQA}} & \multicolumn{4}{c}{\cellcolor{gray!15}\textbf{NQ}} \\
Method & Acc$\uparrow$ & CF & KF & MFS$\uparrow$ & Acc$\uparrow$ & CF & KF & MFS$\uparrow$ & Acc$\uparrow$ & CF & KF & MFS$\uparrow$ & Acc$\uparrow$ & CF & KF & MFS$\uparrow$ & Acc$\uparrow$ & CF & KF & MFS$\uparrow$ \\
\midrule
\rowcolor{model1}
\multicolumn{21}{c}{\textbf{Qwen-2.5-7B}} \\
\midrule
Closed-Book & 30.1 & 4.1 & 100.0 & 52.0 & 14.6 & 1.3 & 100.0 & 50.6 & 18.8 & 1.1 & 100.0 & 50.6 & 56.7 & 9.4 & 100.0 & 54.7 & 26.4 & 2.6 & 100.0 & 51.3 \\
Vanilla RAG & 52.1 & 100.0 & 6.7 & 53.4 & 44.5 & 100.0 & 7.0 & 53.5 & 66.5 & 100.0 & 0.0 & 50.0 & 52.9 & 100.0 & 3.9 & 52.0 & 47.5 & 100.0 & 0.0 & 50.0 \\
\midrule
Internal-Eval & 51.6 & 83.3 & 55.6 & 69.5 & 38.2 & 72.1 & 63.0 & 67.6 & 59.4 & 78.9 & 43.4 & 61.1 & 65.4 & 71.7 & 84.3 & 78.0 & 43.4 & 70.5 & 58.1 & 64.3 \\
Context-Eval & 44.8 & 53.6 & 89.2 & 71.4 & 31.0 & 55.9 & 50.0 & 53.0 & 60.7 & 82.5 & 42.7 & 62.6 & 59.2 & 79.2 & 51.0 & 65.1 & 44.6 & 75.6 & 48.4 & 62.0 \\
Implicit-SCR & \cellcolor{rankfirst}\textbf{59.5} & 90.4 & 59.9 & \cellcolor{ranksecond}{75.2} & \cellcolor{ranksecond}{47.8} & 94.5 & 36.0 & 65.3 & \cellcolor{ranksecond}{68.6} & 93.1 & 13.3 & 53.2 & 67.1 & 100.0 & 52.9 & 76.5 & \cellcolor{ranksecond}{52.5} & 93.6 & 35.5 & 64.5 \\
Explicit-SCR & 40.6 & 57.3 & 44.7 & 51.0 & 34.8 & 70.7 & 30.0 & 50.3 & 58.6 & 81.2 & 28.7 & 54.9 & 50.0 & 62.3 & 51.0 & 56.6 & 41.3 & 65.4 & 54.8 & 60.1 \\
\midrule
TPC & 52.8 & 92.7 & 43.9 & 68.3 & 44.3 & 96.0 & 27.0 & 61.5 & 66.4 & 94.7 & 25.2 & 59.9 & 58.8 & 92.5 & 33.3 & 62.9 & 50.0 & 97.4 & 29.0 & 63.2 \\
TACS-LR & 45.4 & 67.1 & 35.1 & 51.1 & 41.9 & 84.0 & 13.0 & 48.5 & 65.8 & 90.7 & 13.3 & 52.0 & 55.4 & 94.3 & 19.6 & 57.0 & 47.9 & 89.7 & 19.4 & 54.5 \\
\midrule
R-Tuning & 38.0 & 18.3 & 78.2 & 48.3 & 15.9 & 4.4 & 70.9 & 37.7 & 18.5 & 5.1 & 55.0 & 30.1 & 67.5 & 34.7 & 93.1 & 63.9 & 33.5 & 16.7 & 81.8 & 49.3 \\
CR-DPO & 51.6 & 83.3 & 57.9 & 70.6 & 47.8 & 98.5 & 61.0 & \cellcolor{ranksecond}{79.8} & 62.8 & 85.7 & 41.3 & \cellcolor{ranksecond}{63.5} & \cellcolor{rankfirst}\textbf{72.5} & 92.5 & 86.3 & \cellcolor{rankfirst}\textbf{89.4} & 49.6 & 78.2 & 71.0 & \cellcolor{ranksecond}{74.6} \\
\midrule
\textbf{SABER (Ours)} & \cellcolor{ranksecond}{57.0} & 91.8 & 90.1 & \cellcolor{rankfirst}\textbf{90.9} & \cellcolor{rankfirst}\textbf{49.6} & 99.3 & 90.0 & \cellcolor{rankfirst}\textbf{94.6} & \cellcolor{rankfirst}\textbf{73.5} & 96.4 & 75.5 & \cellcolor{rankfirst}\textbf{86.0} & \cellcolor{ranksecond}{71.2} & 86.8 & 88.2 & \cellcolor{ranksecond}{87.5} & \cellcolor{rankfirst}\textbf{53.7} & 92.3 & 71.0 & \cellcolor{rankfirst}\textbf{81.6} \\
\midrule
\rowcolor{model2}
\multicolumn{21}{c}{\textbf{Qwen-2.5-3B}} \\
\midrule
Closed-Book & 24.8 & 4.7 & 100.0 & 52.3 & 9.1 & 1.1 & 100.0 & 50.6 & 13.6 & 0.3 & 100.0 & 50.1 & 47.5 & 8.2 & 100.0 & 54.1 & 27.3 & 2.7 & 100.0 & 51.4 \\
Vanilla RAG & 52.7 & 100.0 & 9.1 & 54.5 & 42.1 & 100.0 & 7.0 & 53.5 & 65.9 & 100.0 & 0.9 & 50.4 & 51.7 & 100.0 & 8.7 & 54.3 & 45.5 & 100.0 & 6.5 & 53.2 \\
\midrule
Internal-Eval & 45.8 & 74.1 & 40.2 & 57.1 & 30.6 & 64.9 & 57.9 & 61.4 & 41.8 & 53.5 & 51.7 & 52.6 & \cellcolor{ranksecond}{63.3} & 86.9 & 78.3 & \cellcolor{ranksecond}{82.6} & 41.7 & 58.1 & 83.9 & \cellcolor{ranksecond}{71.0} \\
Context-Eval & 30.8 & 22.0 & 96.7 & 59.4 & 12.3 & 13.8 & 70.2 & 42.0 & 39.6 & 47.3 & 70.7 & \cellcolor{ranksecond}{59.0} & 49.6 & 21.3 & 95.7 & 58.5 & 31.0 & 29.7 & 71.0 & 50.3 \\
Implicit-SCR & \cellcolor{ranksecond}{53.2} & 86.3 & 33.0 & 59.6 & \cellcolor{ranksecond}{43.4} & 92.9 & 19.3 & 56.1 & \cellcolor{ranksecond}{66.4} & 93.2 & 5.2 & 49.2 & 53.3 & 93.4 & 26.1 & 59.8 & \cellcolor{ranksecond}{48.3} & 95.9 & 19.4 & 57.7 \\
Explicit-SCR & 39.8 & 69.2 & 35.5 & 52.4 & 31.8 & 78.4 & 14.0 & 46.2 & 47.1 & 72.1 & 7.8 & 39.9 & 39.6 & 83.6 & 17.4 & 50.5 & 31.4 & 71.6 & 9.7 & 40.6 \\
\midrule
TPC & 52.4 & 93.3 & 39.1 & \cellcolor{ranksecond}{66.2} & 41.6 & 96.7 & 40.4 & \cellcolor{ranksecond}{68.5} & 63.0 & 90.9 & 24.1 & 57.5 & 55.4 & 95.1 & 30.4 & 62.8 & 46.7 & 98.6 & 25.8 & 62.2 \\
TACS-LR & 45.9 & 81.3 & 17.4 & 49.4 & 39.5 & 90.9 & 3.5 & 47.2 & 62.3 & 89.0 & 0.0 & 44.5 & 50.4 & 90.2 & 15.2 & 52.7 & 45.0 & 94.6 & 12.9 & 53.7 \\
\midrule
R-Tuning & 41.8 & 31.2 & 76.0 & 53.6 & 10.4 & 6.4 & 46.2 & 26.3 & 8.6 & 1.9 & 40.9 & 21.4 & 56.2 & 35.5 & 80.8 & 58.2 & 28.1 & 11.7 & 78.8 & 45.3 \\
CR-DPO & 48.3 & 85.4 & 34.8 & 60.1 & 38.6 & 92.2 & 31.6 & 61.9 & 48.8 & 69.1 & 32.8 & 50.9 & 54.2 & 83.6 & 60.9 & 72.2 & 36.0 & 73.0 & 22.6 & 47.8 \\
\midrule
\textbf{SABER (Ours)} & \cellcolor{rankfirst}\textbf{55.9} & 94.0 & 76.4 & \cellcolor{rankfirst}\textbf{85.2} & \cellcolor{rankfirst}\textbf{44.0} & 98.0 & 75.4 & \cellcolor{rankfirst}\textbf{86.7} & \cellcolor{rankfirst}\textbf{71.6} & 97.6 & 73.3 & \cellcolor{rankfirst}\textbf{85.4} & \cellcolor{rankfirst}\textbf{63.7} & 85.2 & 87.0 & \cellcolor{rankfirst}\textbf{86.1} & \cellcolor{rankfirst}\textbf{50.0} & 97.3 & 45.2 & \cellcolor{rankfirst}\textbf{71.2} \\
\bottomrule
\end{tabular}}
\vspace{-3mm}
\end{table*}
\subsection{Selective Answering with Abstention}
\label{sec:abstention}

We further evaluate whether SABER and prompt-based baselines can refuse to answer when neither PK nor CK is reliable. For a fair comparison, we augment each prompt-based baseline with an explicit instruction to respond \emph{I don't know} as an option. SABER abstains when both probe scores fall below a shared threshold \(\tau=0.5\), the natural midpoint of the $[0,1]$ belief scale. This default is used across all backbones, with no per-backbone tuning. Following the selective-prediction framework of \citet{geifman2017}, each answered instance receives \(+1\) if correct and \(-1\) if wrong, while abstentions receive \(0\). We report four metrics in percent. \textbf{Score} aggregates this utility and measures the answer-abstain trade-off. \textbf{Coverage} (Cov) is the fraction of questions answered. \textbf{Risk on answered} (\(R_C\)) is the error rate among answered questions. \textbf{Abstention F1} (\(F_1\)) measures how well abstentions match the gold abstention set, namely instances where neither source is correct.

\textbf{Main comparison.} Table~\ref{tab:abstain-main} reports results across four backbones. We highlight three observations.

\textbf{(1) SABER achieves the best abstention-aware performance on every backbone.} It obtains the highest Score, lowest Risk on answered, and highest abstention F1, with Implicit-SCR consistently second on Score. The stable gap suggests that prompt-based abstention can only elicit what the model verbalizes about uncertainty, while reliable refusal requires access to internal belief states.

\textbf{(2) SABER's Score advantage grows as backbone size decreases.} The gain over the strongest baseline increases from \(+18.5\%\) on Llama-3.1-8B to \(+24.6\%\) on Qwen-2.5-7B, \(+27.1\%\) on Llama-3.2-3B, and \(+32.0\%\) on Qwen-2.5-3B. This pattern suggests that prompt-based \emph{I don't know} behavior depends on instruction-following capacity and degrades on smaller models, whereas SABER's hidden-state-based decision avoids this dependency.

\textbf{(3) Judgement-style baselines fail structurally under abstention.} Context-Eval, Explicit-SCR, and Internal-Eval answer most questions and achieve much lower abstention F1 than SABER. Their verbal-judgement prompts still frame the task as choosing whether to trust PK or CK, leaving no natural role for abstention. Reliable refusal therefore requires belief estimation, not an instruction.

\textbf{Risk-coverage analysis.} The main comparison uses one operating point. To test whether SABER's advantage is robust, we sweep its abstention threshold \(\tau\) over \([0.05, 0.95]\) in steps of \(0.05\) and trace the full risk-coverage curve. Each baseline yields a single Coverage-Risk point determined by its prompt template. Figure~\ref{fig:abstain-rc-curve} visualizes this comparison, and full sweep numbers are reported in Appendix~\ref{sec:appendix-tau-sweep}.

\textbf{(4) SABER's risk-coverage curve Pareto-dominates every prompt-based baseline on all four backbones.} At Implicit-SCR's Coverage, SABER reduces Risk on answered by \(19\%\) to \(53\%\) relative across the four backbones, with the largest reduction on Llama-3.1-8B. Prompt-based methods commit to one operating point, whereas SABER's continuous threshold \(\tau\) provides a deployment-relevant knob for trading willingness to answer against answer risk without retraining.

\textbf{(5) SABER's Score remains stable over \(\tau \in [0.40, 0.55]\).} The per-backbone fluctuation is at most \(0.009\), showing that the default \(\tau=0.5\) does not require per-backbone recalibration. This stability is consistent with the view that confident instances concentrate away from the decision boundary, making Score insensitive to small threshold changes.

\begin{table*}[t]
\centering
\vspace{-8mm}
\scriptsize
\setlength{\tabcolsep}{2.5pt}
\caption{Selective answering results across four backbones over the whole benchmark ($\tau = 0.5$). We report Score ($\uparrow$), Coverage (Cov), Risk on answered (R$_C$, $\downarrow$), and abstention F1 (F$_1$, $\uparrow$), all as percentages. For Score, R$_C$, and F$_1$, within each backbone block the best value is highlighted with a {\setlength{\fboxsep}{1pt}\colorbox{rankfirst}{\textbf{blue}}} background and the second-best with a {\setlength{\fboxsep}{1pt}\colorbox{ranksecond}{green}} background; Coverage receives no markings.}
\label{tab:abstain-main}
\resizebox{\textwidth}{!}{%
\begin{tabular}{l|cccc|cccc|cccc|cccc}
\toprule
 & \multicolumn{4}{c|}{\cellcolor{gray!15}\textbf{Llama-3.1-8B}} & \multicolumn{4}{c|}{\cellcolor{gray!15}\textbf{Qwen-2.5-7B}} & \multicolumn{4}{c|}{\cellcolor{gray!15}\textbf{Llama-3.2-3B}} & \multicolumn{4}{c}{\cellcolor{gray!15}\textbf{Qwen-2.5-3B}} \\
Method & Score$\uparrow$ & Cov & R$_C\!\downarrow$ & F$_1\!\uparrow$ & Score$\uparrow$ & Cov & R$_C\!\downarrow$ & F$_1\!\uparrow$ & Score$\uparrow$ & Cov & R$_C\!\downarrow$ & F$_1\!\uparrow$ & Score$\uparrow$ & Cov & R$_C\!\downarrow$ & F$_1\!\uparrow$ \\
\midrule
Closed-Book   & $-$3.0 & 50.6 & 53.0 & 50.1 & 1.7 & 23.2 & 46.4 & \cellcolor{ranksecond}{58.0} & $-$5.0 & 37.0 & 56.8 & \cellcolor{ranksecond}{57.3} & $-$0.3 & 6.3 & 52.4 & \cellcolor{ranksecond}{62.3} \\
Vanilla RAG   & 19.8 & 72.9 & 36.4 & 49.6 & 11.1 & 87.1 & 43.6 & 27.9 & 8.2 & 79.4 & 44.9 & 43.1 & 6.2 & 79.5 & 46.1 & 35.8 \\
\midrule
Internal-Eval & 0.5 & 89.2 & 49.7 & 20.0 & 0.8 & 45.3 & 49.2 & 50.4 & $-$11.9 & 72.5 & 58.2 & 38.3 & $-$6.8 & 68.4 & 54.9 & 40.8 \\
Context-Eval  & $-$7.6 & 99.2 & 53.8 & 3.3 & $-$1.8 & 77.6 & 51.1 & 27.3 & $-$34.6 & 100.0 & 67.3 & 0.1 & $-$33.5 & 99.3 & 66.9 & 1.9 \\
Implicit-SCR  & \cellcolor{ranksecond}{25.4} & 59.8 & \cellcolor{ranksecond}{28.8} & \cellcolor{ranksecond}{56.2} & \cellcolor{ranksecond}{18.8} & 81.6 & 38.5 & 39.8 & \cellcolor{ranksecond}{11.4} & 70.7 & \cellcolor{ranksecond}{41.9} & 51.0 & \cellcolor{ranksecond}{7.6} & 73.0 & \cellcolor{ranksecond}{44.8} & 42.0 \\
Explicit-SCR  & $-$8.7 & 97.0 & 54.5 & 8.3 & 11.4 & 43.7 & \cellcolor{ranksecond}{36.9} & 55.4 & $-$34.2 & 97.6 & 67.5 & 4.9 & $-$5.9 & 43.2 & 56.8 & 54.5 \\
\midrule
TACS-LR       & 13.6 & 64.4 & 39.5 & 49.9 & 8.7 & 68.3 & 43.7 & 42.8 & 5.4 & 68.7 & 46.1 & 47.5 & 5.1 & 71.2 & 46.4 & 40.6 \\
\midrule
\textbf{SABER (Ours)} & \cellcolor{rankfirst}\textbf{43.9} & 65.0 & \cellcolor{rankfirst}\textbf{16.2} & \cellcolor{rankfirst}\textbf{72.5} & \cellcolor{rankfirst}\textbf{43.4} & 58.5 & \cellcolor{rankfirst}\textbf{12.9} & \cellcolor{rankfirst}\textbf{78.0} & \cellcolor{rankfirst}\textbf{38.5} & 58.6 & \cellcolor{rankfirst}\textbf{17.2} & \cellcolor{rankfirst}\textbf{76.8} & \cellcolor{rankfirst}\textbf{39.6} & 60.0 & \cellcolor{rankfirst}\textbf{17.0} & \cellcolor{rankfirst}\textbf{76.6} \\
\bottomrule
\end{tabular}}
\vspace{-3mm}
\end{table*}

\begin{figure}[t]
\centering
\includegraphics[width=0.8\linewidth]{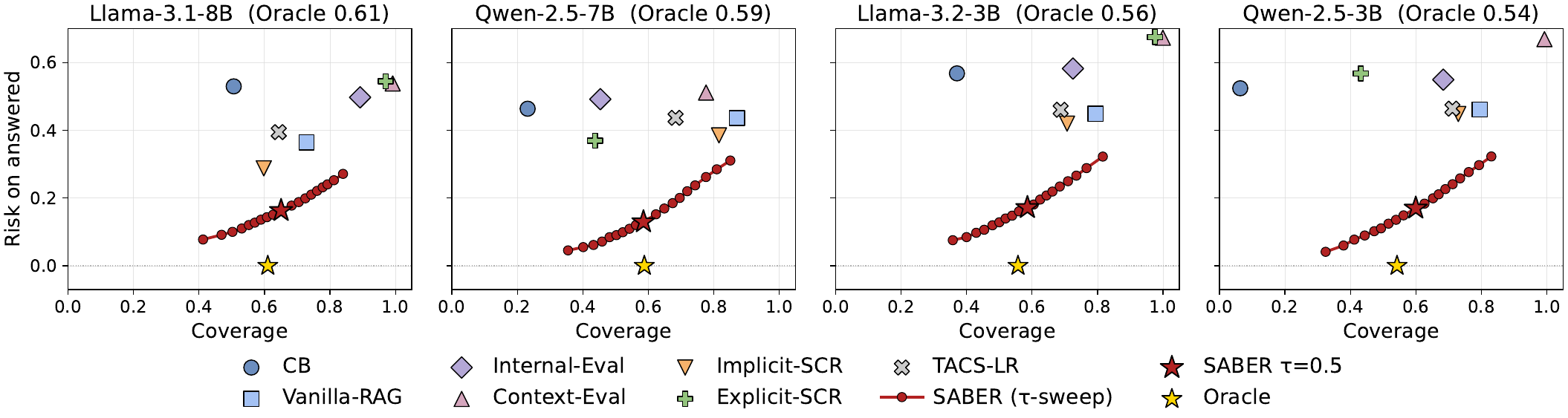}
\caption{Risk-coverage curves. Each panel corresponds to one LLM backbone, and lower-left indicates better performance. SABER (red curve) sweeps its abstention threshold $\tau$ over $[0.05, 0.95]$, while each prompt-based baseline yields a single (Coverage, Risk on answered) point. The dark red star marks SABER at the default $\tau = 0.5$, and the gold star marks the Oracle.}
\label{fig:abstain-rc-curve}
\vspace{-4mm}
\end{figure}

\begin{figure}[t]
\centering
\vspace{-6mm}
\includegraphics[width=0.84\linewidth]{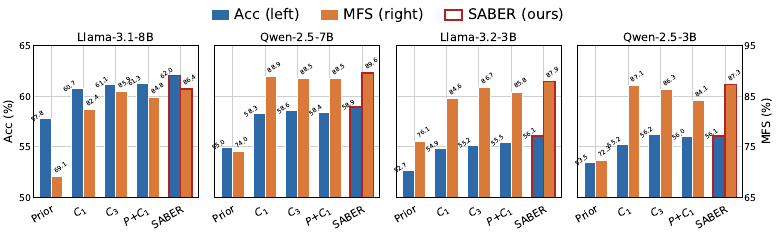}
\vspace{-3mm}
\caption{Ablation study with four variants across four LLMs over the whole benchmark. Each panel reports answer accuracy and MFS. The SABER bars are outlined in red.}
\vspace{-2mm}
\label{fig:ablation-bars}
\end{figure}

\begin{figure}[t]
\centering
\includegraphics[width=0.65\linewidth]{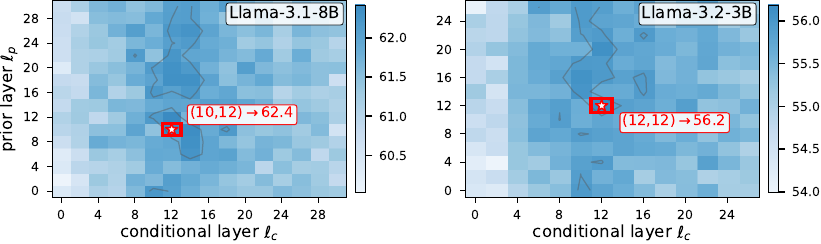}
\vspace{-2mm}
\caption{Layer sensitivity of SABER on the Llama family. Each panel shows end-to-end answer accuracy (\%) over the grid of $(\ell_p, \ell_c)$ at stride~$2$. Darker cells denote higher accuracy, and the grey contour traces the 90th-percentile region. The selected layer pair is marked by the red square. The Qwen counterpart is in Appendix~\ref{sec:appendix-qwen-heatmap}.}
\label{fig:grid2d-acc-llama}
\vspace{-2mm}
\end{figure}

\subsection{Ablation Studies}
\label{sec:ablation}

To assess which components drive SABER's performance, we compare the full method with four reduced variants: a prior-only baseline (Prior, using only the query-side hidden state), two conditional-only variants (\textbf{\(C_1\)} and \textbf{\(C_3\)}, using one or three mean-pooled traces), and a single-trace joint variant (\textbf{\(P{+}C_1\)}). Our full method, SABER, is the joint variant with \(K{=}3\), equivalent to \(P{+}C_3\). Figure~\ref{fig:ablation-bars} reports Acc and MFS, with SABER bars outlined in red. We highlight three observations.

\textbf{(1) Each component contributes to SABER's overall performance.} Every reduced variant trails SABER in both Acc and MFS on all four backbones. Among the single-component removals, dropping multi-trace averaging from the joint configuration causes the largest MFS reduction, up to \(3.2\%\) on Qwen-2.5-3B. This indicates that the components capture complementary signals.

\textbf{(2) The source-conditional reasoning representation drives most of the gain.} Conditional-only variants \(C_1\) and \(C_3\) improve over the prior-only baseline by an average of \(+2.8\%\) Acc and \(+13.4\%\) MFS across the four backbones. This indicates that the post-reasoning, source-conditioned state carries most reliability information, while the query-only prior is useful but insufficient alone.

\textbf{(3) Multi-trace averaging benefits all backbones, with the largest gain on the smallest model.} For example, on Qwen-2.5-3B, increasing \(K\) from \(1\) to \(3\) in the joint variant, from \(P{+}C_1\) to SABER, raises MFS from \(84.1\%\) to \(87.3\%\) (\(+3.2\%\)), the largest gain across the four backbones. This suggests that mean-pooling multiple conditional traces reduces noise from single samples, and the noise reduction matters most when the underlying LLM backbone is smaller.

\subsection{Layer Sensitivity}
\label{sec:layer-sensitivity}

We examine SABER's sensitivity to the prior layer \(\ell_p\) and conditional layer \(\ell_c\). Figure~\ref{fig:grid2d-acc-llama} sweeps both layers at stride~\(2\) over the two Llama backbones. The best layer pair lies within a broad high-accuracy plateau in the mid-to-late depth range. For both Llama-3.1-8B and Llama-3.2-3B, the 90th-percentile region covers more than one third of the grid, and the selected layer pair (red square) lies inside this region rather than at an isolated peak. The selected layer pair on each Llama backbone falls between roughly 40\% and 70\% of the total depth \(L\), sitting within the mid-to-late depth band that prior probing work identifies as semantically integrated~\citep{azaria2023internal,burns2022discovering}. Thus, SABER is not sensitive to fine-grained layer tuning and uses depths known to be informative for internal belief estimation.

\vspace{-2mm}
\section{Related Work}
\vspace{-2mm}
\textbf{Knowledge Conflict in RAG.} Prior benchmarks formalize the CK-PK relation as supportive, contradictory, or irrelevant \citep{mallen2023nottrust,xie2024adaptive,wu2025clasheval,bi2024confiqa,su2024conflictbank,longpre2021entity,hou2024wikicontradict,xu2024kcsurvey}. This data-level view helps reveal PK-CK mismatch, but overlooks each model's actual answers and their correctness. Along this line, some works infer conflicts from token likelihoods or distributional uncertainty \citep{wu2025clasheval,shi2024cad}, some steer generation through decoding interventions or verbal source judgments \citep{shi2024cad,huang2025trust,wang2025adacad}, and others revise retrieved evidence or adaptively decide when retrieval is needed \citep{yu2024tacs,zhang2025faithfulrag,gao2025clear,asai2024selfrag,jeong2024adaptiverag,su2024dragin,yao2025seakr}. These methods improve source coordination, but without correctness estimation, they may still force unsupported answers when abstention is warranted. In contrast, we label PK- and CK-path correctness against ground truth and enable calibrated abstention when neither path is reliable.

\textbf{Self-Awareness in LLMs.} A growing body of work investigates whether LLMs can recognize the limits of their own knowledge and reasoning. On the knowledge side, LLMs expose partial self-knowledge through calibrated confidence, verbalized uncertainty, answerability estimation, and hidden-state probes of latent truthfulness or hallucination risk \citep{kadavath2022language,lin2022teaching,chen2026query,burns2022discovering,azaria2023internal,yin2023dontknow,chen2024inside}. On the reasoning side, recent work examines whether models can evaluate intermediate reasoning states or anticipate response-level failures from internal representations \citep{ni2026reprobe,ghasemabadi2026gnosis}. These studies typically extract a single self-awareness signal for post-hoc diagnosis, such as calibration, hallucination detection, or selective prediction on the model's own answer. SABER instead combines a knowledge-boundary self-prior with reasoning-reliability self-evaluation into paired PK- and CK-side beliefs, and uses them at inference time to choose PK, choose CK, or abstain when both paths fail \citep{geifman2017,zhang2024rtuning,wen2025abstention}.

\vspace{-2mm}

\section{Conclusion}
\label{sec:conclusion}

\vspace{-2mm}
We presented SABER, a Self-Aware Belief Estimator that resolves RAG knowledge conflicts by estimating PK- and CK-side correctness beliefs from frozen LLM hidden states. We also constructed a model-specific benchmark labeling PK-only and CK-conditioned answer correctness across four Llama and Qwen backbones. SABER improves end-to-end accuracy and faithfulness over strong baselines and provides a calibrated abstention signal that Pareto-dominates prompt-based abstainers on every backbone. Ablations and layer-sensitivity studies show that these gains come from combining a self-prior with source-conditional reasoning representations and remain robust across a broad mid-to-late layer plateau. More broadly, our results suggest that self-awareness is a promising foundation for faithful RAG that can recognize unsupported cases and abstain rather than force an answer.

\bibliographystyle{plainnat}
\bibliography{references}
\newpage

\appendix
\clearpage

{\centering\bfseries\Large Appendix\par}
\vspace{0.6em}

\section{Implementation Details}
\label{sec:appendix-implementation}

\subsection{Probe architecture and training}
\label{sec:appendix-probe}

The two side-specific reliability predictors $f_{\mathrm{pred}, \mathrm{PK}}$ and $f_{\mathrm{pred}, \mathrm{CK}}$ share the same architecture and training recipe, summarised in Table~\ref{tab:appendix-probe}. The LLM backbone is kept frozen throughout.

\begin{table}[!ht]
\centering
\small
\caption{Probe architecture and training hyperparameters for the two reliability predictors $f_{\mathrm{pred}, \mathrm{PK}}$ and $f_{\mathrm{pred}, \mathrm{CK}}$, which share the same architecture and training recipe.}
\label{tab:appendix-probe}
\begin{tabular}{ll}
\toprule
\textbf{Component} & \textbf{Setting} \\
\midrule
Hidden sizes & $(256, 128)$ \\
Activation & ReLU \\
Dropout & $0.2$ \\
Input normalisation & Per-feature standardisation (fit on train) \\
Optimiser & AdamW \\
Learning rate & $1{\times}10^{-3}$ \\
Weight decay & $1{\times}10^{-4}$ \\
Batch size & $256$ \\
Loss & Binary cross-entropy with logits \\
Max epochs & $50$ \\
Early stopping & On validation AUROC, patience $8$ \\
Seed & $42$ \\
\bottomrule
\end{tabular}
\end{table}

\subsection{Multi-trace generation}
\label{sec:appendix-generation}

We sample $K{=}3$ reasoning traces per side using the parameters in Table~\ref{tab:appendix-generation}. The same parameters are applied to both PK-side and CK-side trace generation across all four backbones.

\begin{table}[!ht]
\centering
\small
\caption{Sampling parameters for multi-trace reasoning generation. Each trace is sampled independently with a different seed, and the same parameters are used on both PK and CK sides across all four backbones.}
\label{tab:appendix-generation}
\begin{tabular}{ll}
\toprule
\textbf{Parameter} & \textbf{Value} \\
\midrule
Number of traces $K$ & $3$ \\
Decoding & Stochastic sampling \\
Temperature & $0.9$ \\
Top-$p$ & $1.0$ \\
Repetition penalty & $1.05$ \\
Max new tokens & $256$ \\
Per-trace seeds & $\{0, 1, 2\}$ \\
\bottomrule
\end{tabular}
\end{table}

\subsection{Metric definitions}
\label{sec:appendix-metrics}

All metrics are computed on the same test set of $N$ instances. Each instance is mapped, by alias-matching the closed-book PK answer and the context-grounded CK answer against the gold, to one of four reliability cells $C_{00}, C_{01}, C_{10}, C_{11}$, where $C_{ab}$ collects the instances on which PK is correct iff $a = 1$ and CK is correct iff $b = 1$.

\paragraph{Main-results metrics (\S\ref{sec:main-results}; Tables~\ref{tab:main-llama},~\ref{tab:main-qwen},~\ref{tab:appendix-main-llama-full},~\ref{tab:appendix-main-qwen-full}).} For each instance $i$, let $c_i \in \{0, 1\}$ indicate whether the method's answer alias-matches the gold. The four main-results metrics describe a method along two axes: how often it produces the right answer overall, and how it handles the two single-source cells where the right answer is reachable from only one of \{PK, CK\}.

\begin{itemize}
\item \textbf{Acc (Answer accuracy).} The fraction of instances on which the method's answer alias-matches the gold. This is the bottom-line correctness rate; it does not distinguish how the method arrived at the answer or which cell the instance came from.
\begin{equation}
\mathrm{Acc} \;=\; \frac{1}{N}\sum_{i=1}^{N} c_i.
\end{equation}

\item \textbf{CF (Context Faithfulness).} Accuracy restricted to the cell $C_{01}$, where only the context carries the correct answer. CF asks: when the model would be wrong from its parametric knowledge alone but the retrieved context contains the right answer, does the method actually follow the context? A high CF indicates that the method effectively absorbs information from a trustworthy context.
\begin{equation}
\mathrm{CF} \;=\; \frac{1}{|C_{01}|}\sum_{i \in C_{01}} c_i.
\end{equation}

\item \textbf{KF (Knowledge Faithfulness).} The mirror of CF, restricted to $C_{10}$, where only parametric knowledge is correct and the retrieved context is misleading. KF asks: when the context would lead the model astray, does the method preserve its internally-correct PK answer? A high KF indicates that the method is robust to a misleading context.
\begin{equation}
\mathrm{KF} \;=\; \frac{1}{|C_{10}|}\sum_{i \in C_{10}} c_i.
\end{equation}

\item \textbf{MFS (Macro Faithfulness Score).} The unweighted mean of CF and KF. MFS balances the two failure modes, over-trusting context and over-trusting memory, into a single number; a method that inflates Acc by collapsing onto one source will register near-zero on either CF or KF and so drag MFS down. We treat MFS as the primary single-number measure of conflict-resolution quality.
\begin{equation}
\mathrm{MFS} \;=\; \tfrac{1}{2}\,\big(\mathrm{CF} + \mathrm{KF}\big).
\end{equation}
\end{itemize}

\paragraph{Selective-answering metrics (\S\ref{sec:abstention}; Table~\ref{tab:abstain-main}).} Each instance falls into exactly one of three outcomes -- answered correctly, answered wrongly, or abstained -- with counts $N_+, N_-, N_0$ satisfying $N_+ + N_- + N_0 = N$. Let $\mathcal{A}$ denote the set of instances on which the method abstained, and let $G = C_{00}$ denote the gold-abstain set (instances on which neither PK nor CK is correct).

\begin{itemize}
\item \textbf{Score.} The mean per-instance utility under the selective-prediction scheme of \citet{geifman2017}: a correct answer is worth $+1$, a wrong answer $-1$, and an abstention $0$. Equivalently, Score is the gap between the number of correct and wrong answers normalised by the test-set size. Concretely, define the per-instance utility
\begin{equation}
u_i \;=\;
\begin{cases}
+1 & \text{if instance } i \text{ is answered correctly,}\\
-1 & \text{if instance } i \text{ is answered wrongly,}\\
\phantom{+}0 & \text{if instance } i \text{ is abstained.}
\end{cases}
\end{equation}
The aggregate Score is then
\begin{equation}
\mathrm{Score} \;=\; \frac{1}{N}\sum_{i=1}^{N} u_i \;=\; \frac{N_+ - N_-}{N}.
\end{equation}
Score rewards a method only when correct answers outweigh wrong ones; because abstaining contributes zero, a method that refuses everywhere collapses to $\mathrm{Score} = 0$ rather than going negative, which prevents trivial gaming by silence.

\item \textbf{Cov (Coverage).} The fraction of instances on which the method commits to an answer. Coverage measures \emph{willingness to answer}; we report it as a diagnostic axis rather than a goal, since it can be raised arbitrarily by simply refusing to abstain.
\begin{equation}
\mathrm{Cov} \;=\; \frac{N_+ + N_-}{N}.
\end{equation}

\item \textbf{$R_C$ (Risk on answered).} The error rate among the answered instances, ignoring abstentions. $R_C$ measures the \emph{conditional quality} of the answers a method does commit to; reporting it together with Cov exposes the trade-off between answering more and answering reliably.
\begin{equation}
R_C \;=\; \frac{N_-}{N_+ + N_-}.
\end{equation}

\item \textbf{$F_1$ (Abstention F1).} The F1 of the abstain decisions against the gold-abstain set $G = C_{00}$, treating abstain as the positive class. $F_1$ measures how \emph{targeted} the refusals are: a method that abstains on everything reaches near-perfect recall but very low precision, while a method that never abstains scores zero on both, so $F_1$ penalises both blanket caution and reckless answering.
\begin{equation}
F_1 \;=\; \frac{2\,|\mathcal{A} \cap G|}{|\mathcal{A}| + |G|}.
\end{equation}
\end{itemize}

Together, Score, Cov, $R_C$, and $F_1$ span three complementary axes, including net utility (Score), the answer-versus-refuse trade-off (Cov, $R_C$), and abstain targeting ($F_1$), that resist gaming by any single failure mode.

\section{Datasets}
\label{sec:appendix-datasets}

\subsection{Source datasets}
\label{sec:appendix-datasets-used}

Our benchmark draws from five publicly released QA datasets, each exposing a controlled form of PK--CK conflict.

\begin{itemize}
\item \textbf{ConFiQA} \citep{bi2024confiqa}: A QA benchmark where the retrieved context is modified by counterfactual substitution to assert an answer that contradicts the ground truth. We use all three sub-tasks (QA, MR, MC), which differ in question structure but share the same conflict-construction pipeline.
\item \textbf{ConflictQA-PopQA} \citep{xie2024adaptive}: Built from PopQA queries on which the LLM has previously given a wrong answer; the retrieved passage is constructed to support the model's incorrect prediction. This directly probes the model's tendency to trust a misleading document over its own (correct) parametric knowledge.
\item \textbf{ConflictBank} \citep{su2024conflictbank}: A synthesised benchmark of factual conflicts covering multiple categories such as entity substitution, temporal mismatch, and attribute alteration. We use a pilot subset that spans the main categories.
\item \textbf{TriviaQA}: We use the adversarial-context subset prepared by \citet{huang2025trust}, in which the original supporting passage is perturbed to assert a wrong answer, applying the same construction style as ConflictQA-PopQA. It tests robustness to misleading context on questions about widely-known entities.
\item \textbf{NQ}: We use the adversarial-context subset of Natural Questions prepared by \citet{huang2025trust}, applying the same construction as the TriviaQA subset above on naturally-occurring user queries. It complements TriviaQA by covering a different query distribution.
\end{itemize}

\subsection{Benchmark statistics and processing}
\label{sec:appendix-dataset-processing}

\begin{table}[!ht]
\centering
\small
\caption{Constructed benchmark sizes per backbone (after processing). The same partition structure is used for every backbone, and per-backbone row counts differ slightly because instances that fail end-to-end inference are dropped per backbone.}
\label{tab:appendix-dataset-stats}
\begin{tabular}{lrrrr}
\toprule
\textbf{Sub-dataset} & $N_{\mathrm{train}}$ & $N_{\mathrm{val}}$ & $N_{\mathrm{test}}$ & \textbf{Total} \\
\midrule
ConFiQA-QA & 9{,}600 & 1{,}200 & 1{,}200 & 12{,}000 \\
ConFiQA-MR & 9{,}600 & 1{,}200 & 1{,}200 & 12{,}000 \\
ConFiQA-MC & 9{,}598 & 1{,}202 & 1{,}200 & 12{,}000 \\
ConflictQA-PopQA & 12{,}834 & 1{,}608 & 1{,}612 & 16{,}054 \\
ConflictBank & 9{,}611 & 1{,}197 & 1{,}192 & 12{,}000 \\
TriviaQA & 1{,}920 & 240 & 240 & 2{,}400 \\
NQ & 1{,}934 & 242 & 242 & 2{,}418 \\
\midrule
\textbf{Total per backbone} & \textbf{55{,}097} & \textbf{6{,}889} & \textbf{6{,}886} & \textbf{68{,}872} \\
\bottomrule
\end{tabular}
\end{table}

\paragraph{Selection and sampling.} We use the publicly released instances in full for ConFiQA, ConflictQA-PopQA, TriviaQA, and NQ. For ConflictBank, which is substantially larger than the other datasets, we sample a subset of approximately $12{,}000$ instances that spans the main conflict categories. The seven datasets are then pooled into a single ground-truth-aligned benchmark that is constructed independently for each backbone.

\paragraph{Labeling answer paths.} For every query-context instance, we run the backbone under both the PK-only and the CK-conditioned answer paths, and label the resulting answers against the gold answer using an alias-aware match cascade $\mathrm{match}(\cdot, \cdot)$. This yields a 4-cell reliability outcome $(y_{\mathrm{PK}}, y_{\mathrm{CK}})$ for the instance, indicating which of the two paths produced a correct answer.

\paragraph{Alias-aware match cascade.} The $\mathrm{match}$ function compares a generated answer string $a$ against the gold answer, or against any of its accepted aliases when the dataset provides a list of equivalents. It first applies Unicode normalisation, casefolding, and punctuation stripping, then tries an exact string match, and finally falls back to a token-level containment check. The side label $y_s$ is set to $1$ if any of these checks succeeds for $a_s$, and to $0$ otherwise.

\paragraph{Filtering and splits.} Instances that the backbone cannot run end-to-end (for example due to context-length truncation) are dropped, which leads to small per-backbone differences in row counts. We then split each dataset by query identifier into a $80\%$\,/\,$10\%$\,/\,$10\%$ train\,/\,validation\,/\,test partition, with the same query never appearing across splits, and aggregate the per-dataset partitions into the final benchmark whose statistics are reported in Table~\ref{tab:appendix-dataset-stats}.

\begin{figure}[!ht]
\centering
\includegraphics[width=\linewidth]{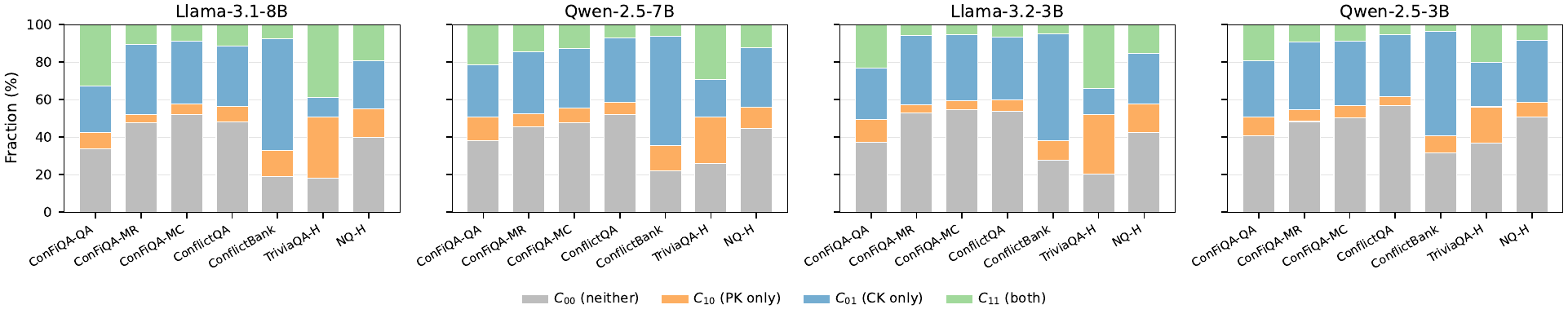}
\caption{4-cell reliability distribution of the constructed benchmark. Each stacked bar shows, for one (backbone, dataset) pair, the fraction of instances falling into the four cells $C_{00}$ (neither side correct), $C_{10}$ (only PK correct), $C_{01}$ (only CK correct), and $C_{11}$ (both correct).}
\label{fig:appendix-cell-dist}
\end{figure}

Figure~\ref{fig:appendix-cell-dist} and Table~\ref{tab:appendix-cell-dist} report the resulting 4-cell distribution per (backbone, dataset). The cell composition varies substantially across datasets and backbones, which gives our predictors a balanced mixture of all four reliability outcomes during training.

\begin{table*}[!ht]
\centering
\small
\setlength{\tabcolsep}{3pt}
\caption{Per-(backbone, dataset) 4-cell distribution as percentages of dataset size, complementing Figure~\ref{fig:appendix-cell-dist}. Rows within each backbone block sum to approximately $100\%$.}
\label{tab:appendix-cell-dist}
\resizebox{\textwidth}{!}{%
\begin{tabular}{l|cccc|cccc|cccc|cccc}
\toprule
 & \multicolumn{4}{c|}{\cellcolor{gray!15}\textbf{Llama-3.1-8B}} & \multicolumn{4}{c|}{\cellcolor{gray!15}\textbf{Qwen-2.5-7B}} & \multicolumn{4}{c|}{\cellcolor{gray!15}\textbf{Llama-3.2-3B}} & \multicolumn{4}{c}{\cellcolor{gray!15}\textbf{Qwen-2.5-3B}} \\
Sub-dataset & $C_{00}$ & $C_{10}$ & $C_{01}$ & $C_{11}$ & $C_{00}$ & $C_{10}$ & $C_{01}$ & $C_{11}$ & $C_{00}$ & $C_{10}$ & $C_{01}$ & $C_{11}$ & $C_{00}$ & $C_{10}$ & $C_{01}$ & $C_{11}$ \\
\midrule
ConFiQA-QA     & 33.9 & 8.7 & 24.6 & 32.8 & 38.3 & 12.7 & 27.6 & 21.5 & 37.4 & 12.1 & 27.4 & 23.0 & 40.9 & 9.6 & 30.3 & 19.1 \\
ConFiQA-MR     & 47.8 & 4.4 & 37.4 & 10.4 & 45.6 & 7.1 & 32.7 & 14.7 & 53.0 & 4.1 & 37.2 & 5.7  & 48.4 & 6.4 & 35.8 & 9.4  \\
ConFiQA-MC     & 52.1 & 5.6 & 33.2 & 9.0  & 47.6 & 8.1 & 31.6 & 12.7 & 54.8 & 4.5 & 35.4 & 5.4  & 50.2 & 6.7 & 34.3 & 8.8  \\
ConflictQA-PopQA & 48.0 & 8.6 & 31.8 & 11.7 & 51.8 & 6.6 & 34.6 & 7.0  & 53.6 & 6.2 & 33.8 & 6.5  & 56.7 & 5.0 & 33.1 & 5.2  \\
ConflictBank   & 19.0 & 14.1 & 59.3 & 7.7  & 21.9 & 13.8 & 58.3 & 6.0  & 27.6 & 10.4 & 57.2 & 4.9  & 31.6 & 9.0 & 55.8 & 3.6  \\
TriviaQA       & 18.0 & 32.7 & 10.3 & 39.0 & 25.8 & 24.9 & 20.2 & 29.1 & 20.2 & 32.0 & 13.6 & 34.3 & 36.7 & 19.5 & 23.6 & 20.2 \\
NQ             & 39.7 & 15.5 & 25.3 & 19.4 & 44.7 & 11.1 & 31.8 & 12.4 & 42.6 & 15.0 & 27.0 & 15.3 & 50.7 & 8.1 & 33.0 & 8.3  \\
\bottomrule
\end{tabular}}
\end{table*}

\section{Baselines}
\label{sec:appendix-baselines}

We compare against ten baselines that span four design philosophies for handling PK--CK conflict: vanilla retrieve-or-not, verbal self-evaluation, probability- or probe-based heuristics, and fine-tuning.

\begin{itemize}
\item \textbf{Closed-Book}: The LLM answers from its parametric memory alone, with no retrieved context. We include it as a non-RAG reference point that characterises the intrinsic knowledge of each backbone before any document is supplied.
\item \textbf{Vanilla RAG} \citep{lewis2020rag}: The standard retrieve-then-read setup, where the LLM answers conditioned on a top-retrieved passage without any further filtering or confidence signal. As the most widely deployed RAG variant, it serves as the natural reference point for any method that adds reliability awareness on top of retrieval.
\item \textbf{Internal-Eval} \citep{huang2025trust}: After producing an internal (PK) answer, the LLM is prompted to judge whether that answer is correct, and the routing decision uses this self-judgement to choose between the PK answer and the retrieved context. It represents a self-confidence-based approach grounded in the model's own assessment of its parametric knowledge.
\item \textbf{Context-Eval} \citep{huang2025trust}: A counterpart to Internal-Eval on the context side: the LLM is asked whether the retrieved passage is reliable enough to be trusted before producing the answer. It isolates the contextual-reliability axis of the routing decision.
\item \textbf{Implicit-SCR} \citep{huang2025trust}: Samples multiple answers under each source path and uses output consistency as an implicit confidence proxy, taking the most consistent answer as the prediction. It is among the strongest prompt-only baselines reported in prior work on knowledge conflict.
\item \textbf{Explicit-SCR} \citep{huang2025trust}: Prompts the LLM to directly compare its PK answer against the CK answer and pick the more reliable one in a single call, the most direct prompting strategy for source arbitration without any auxiliary model or training.
\item \textbf{TPC} \citep{wu2025clasheval}: Follows the answer path whose token-level probability is higher under the LLM's distribution, treating output likelihood as a confidence signal. It represents a probability-based heuristic that operates at decoding time without any extra prompting or training.
\item \textbf{TACS-LR} \citep{yu2024tacs}: Trains a linear probe on the LLM's hidden states to score and filter individual context tokens, retaining only those judged reliable before generating the answer. As a published probe-based method, it tests whether token-level hidden-state filtering alone can cover the routing benefits we aim for.
\item \textbf{CR-DPO} \citep{huang2025trust}: Fine-tunes the LLM backbone using Direct Preference Optimisation on conflict-resolution preference pairs, extending Explicit-SCR with parametric updates. It serves as a fine-tuning-based comparison point that bounds the gain achievable when the backbone itself is allowed to be retrained.
\item \textbf{R-Tuning} \citep{zhang2024rtuning}: Fine-tunes the LLM backbone on an instruction-tuning dataset that explicitly teaches it to respond \emph{I don't know} on questions whose answers it does not reliably know, rather than confabulating one. Although introduced in a non-RAG setting, R-Tuning is a natural training-based comparison point for SABER's calibrated abstention: both aim at honest refusal when the model lacks reliable knowledge, but R-Tuning achieves it through parametric updates while SABER does so through a frozen-backbone hidden-state probe.
\end{itemize}

\section{Prompt Templates}
\label{sec:appendix-prompts}

\subsection{Benchmark construction prompts}
\label{sec:appendix-prompt-construction}

The PK-only and CK-conditioned answer paths used to label the benchmark (Eq.~\ref{eq:answer-paths}) are produced with the two prompts below. They share the same system instruction and differ only in whether the retrieved context is included before the question.

\begin{promptbox}[title={PK answer prompt (closed-book, benchmark construction)}]
\textbf{[System]} Answer the question with a short phrase.\\[2pt]
\textbf{[User]} Question: \{question\}\\
Answer:
\end{promptbox}

\begin{promptbox}[title={CK answer prompt (with retrieved context, benchmark construction)}]
\textbf{[System]} Answer the question with a short phrase.\\[2pt]
\textbf{[User]} Context: \{ck\_text\}\\[2pt]
Question: \{question\}\\
Answer:
\end{promptbox}

\subsection{Self-prior}
\label{sec:appendix-prompt-prior}

The self-prior representation $h_q$ in Eq.~\ref{eq:self-prior} is read from a forward pass on the raw query string alone, with no system message and no instruction template, so that the hidden state reflects the model's pre-context belief without being biased by surface phrasing.

\begin{promptbox}[title=Self-prior prompt]
\{question\}
\end{promptbox}

\subsection{PK-side reasoning generation}
\label{sec:appendix-prompt-pk}

\begin{promptbox}[title=PK reasoning generation prompt]
You will be given a question. Assume your own knowledge is correct and complete for this question. Reason step by step using ONLY your prior knowledge --- do NOT use any external sources. End with one final answer on the last line.\\[2pt]
Question: \{question\}\\[2pt]
Reasoning:
\end{promptbox}

\subsection{CK-side reasoning generation}
\label{sec:appendix-prompt-ck}

\begin{promptbox}[title=CK reasoning generation prompt]
You will be given a question and a document. Assume the document is reliable and contains the correct facts needed to answer the question. Reason step by step using ONLY the information in the document. End with one final answer on the last line.\\[2pt]
Question: \{question\}\\
Document: \{ck\_text\}\\[2pt]
Reasoning:
\end{promptbox}

\subsection{Self-evaluation}
\label{sec:appendix-prompt-eval}

\begin{promptbox}[title=Self-evaluation prompt]
You are evaluating a proposed reasoning for a question. Use your own knowledge to judge whether the reasoning is factually reliable.\\[2pt]
Question: \{question\}\\[2pt]
Proposed reasoning: \{reasoning\}\\[2pt]
Is the reasoning above factually reliable?\\
(A) Yes\\
(B) No\\[2pt]
Answer:
\end{promptbox}

\subsection{Baseline prompts and abstention modifications}
\label{sec:appendix-prompt-baselines}

We adopt the original baseline prompts from their respective sources \citep{huang2025trust, wu2025clasheval, yu2024tacs} and apply only minimal edits to align their input format with our (question, document, gold answer) schema. The modifications described in this subsection are used \emph{only} for the abstention experiment (\S\ref{sec:abstention}); the main-results comparison (\S\ref{sec:main-results}) uses each baseline's original prompt without any abstention suffix. For the abstention experiment, each prompt-only baseline is augmented with a single trailing instruction that allows the model to refuse to answer; the rest of the original prompt, including any in-context examples, is kept verbatim. For instance, the Implicit-SCR system instruction is modified as shown below.

\begin{promptbox}[title=Implicit-SCR system instruction (original \citep{huang2025trust})]
You will be given a question and a document. The document may not be trustworthy. Use your judgment to assess the reliability of the document. Then, based on both your assessment and your own knowledge, provide the best possible answer.
\end{promptbox}

\begin{promptbox}[title=Implicit-SCR system instruction (our abstention variant)]
You will be given a question and a document. The document may not be trustworthy. Use your judgment to assess the reliability of the document. Then, based on both your assessment and your own knowledge, provide the best possible answer. \textbf{If you cannot answer the question accurately based on either your own knowledge or the provided document, respond exactly with: I don't know.}
\end{promptbox}

The same single-sentence abstention suffix is appended to the system instructions of all other prompt-only baselines.

\section{Additional Results}
\label{sec:appendix-additional-results}

\subsection{Per-sub-task breakdown of main results}
\label{sec:appendix-main-full}

Tables~\ref{tab:main-llama} and~\ref{tab:main-qwen} in the main paper report ConFiQA as a single column averaged across its three sub-tasks (QA, MR, MC). Tables~\ref{tab:appendix-main-llama-full} and~\ref{tab:appendix-main-qwen-full} below give the per-sub-task breakdown for those three sub-tasks; numbers on the other four datasets (ConflictQA, ConflictBank, TriviaQA, NQ) are unchanged and can be read directly from the main-paper tables.

\begin{table*}[!ht]
\centering
\small
\setlength{\tabcolsep}{2.5pt}
\caption{Per-sub-task breakdown of ConFiQA on the Llama family. We report Acc ($\uparrow$), CF, KF, and MFS ($\uparrow$), all as percentages, on the three ConFiQA sub-tasks (QA / MR / MC). For Acc and MFS, within each (backbone, sub-task) cell the best value is highlighted with a {\setlength{\fboxsep}{1pt}\colorbox{rankfirst}{\textbf{blue}}} background and the second-best with a {\setlength{\fboxsep}{1pt}\colorbox{ranksecond}{green}} background.}

\label{tab:appendix-main-llama-full}
\resizebox{0.75\textwidth}{!}{%
\begin{tabular}{l|cccc|cccc|cccc}
\toprule
 & \multicolumn{4}{c|}{\cellcolor{gray!15}\textbf{ConFiQA-QA}} & \multicolumn{4}{c|}{\cellcolor{gray!15}\textbf{ConFiQA-MR}} & \multicolumn{4}{c}{\cellcolor{gray!15}\textbf{ConFiQA-MC}} \\
Method & Acc$\uparrow$ & CF & KF & MFS$\uparrow$ & Acc$\uparrow$ & CF & KF & MFS$\uparrow$ & Acc$\uparrow$ & CF & KF & MFS$\uparrow$ \\
\rowcolor{model1}
\midrule
\multicolumn{13}{c}{\textbf{Llama-3.1-8B}} \\
\midrule
Closed-Book   & 51.2 & 17.2 & 100.0 & 58.6 & 22.3 & 3.7 & 100.0 & 51.8 & 20.0 & 5.8 & 100.0 & 52.9 \\
Vanilla RAG   & 65.8 & 100.0 & 21.2 & 60.6 & \cellcolor{rankfirst}\textbf{57.3} & 100.0 & 18.6 & 59.3 & \cellcolor{ranksecond}{47.8} & 100.0 & 10.8 & 55.4 \\
\midrule
Internal-Eval & 60.3 & 56.4 & 79.8 & 68.1 & 40.6 & 54.3 & 45.8 & 50.1 & 35.2 & 55.6 & 52.3 & 53.9 \\
Context-Eval  & 57.4 & 44.0 & 94.9 & 69.5 & 32.8 & 31.3 & 81.4 & 56.3 & 27.6 & 31.2 & 75.4 & 53.3 \\
Implicit-SCR  & \cellcolor{ranksecond}{66.8} & 83.8 & 63.6 & 73.7 & 52.2 & 75.9 & 39.0 & 57.4 & 42.1 & 77.9 & 18.5 & 48.2 \\
Explicit-SCR  & 55.3 & 56.4 & 79.8 & 68.1 & 33.6 & 40.2 & 62.7 & 51.5 & 31.2 & 43.9 & 80.0 & 62.0 \\
\midrule
TPC           & 66.4 & 88.3 & 67.7 & \cellcolor{ranksecond}{78.0} & 52.9 & 85.4 & 47.5 & 66.4 & 45.9 & 86.8 & 61.5 & \cellcolor{ranksecond}{74.2} \\
TACS-LR       & 61.5 & 89.7 & 36.4 & 63.0 & 39.8 & 52.6 & 37.3 & 44.9 & 29.5 & 45.7 & 15.4 & 30.5 \\
\midrule
R-Tuning      & 32.4 & 10.7 & 41.3 & 26.0 & 14.8 & 8.6 & 23.9 & 16.3 & 14.7 & 8.6 & 29.1 & 18.9 \\
CR-DPO        & 58.6 & 51.2 & 87.9 & 69.5 & 46.4 & 63.7 & 74.6 & \cellcolor{ranksecond}{69.1} & 38.6 & 62.4 & 63.1 & 62.8 \\
\midrule
\textbf{SABER (Ours)} & \cellcolor{rankfirst}\textbf{68.2} & 89.0 & 74.7 & \cellcolor{rankfirst}\textbf{81.9} & \cellcolor{ranksecond}{57.0} & 92.8 & 66.1 & \cellcolor{rankfirst}\textbf{79.5} & \cellcolor{rankfirst}\textbf{50.2} & 95.4 & 73.8 & \cellcolor{rankfirst}\textbf{84.6} \\
\rowcolor{model2}
\midrule
\multicolumn{13}{c}{\textbf{Llama-3.2-3B}} \\
\midrule
Closed-Book   & 40.5 & 10.4 & 100.0 & 55.2 & 16.8 & 4.2 & 100.0 & 52.1 & 14.5 & 2.8 & 100.0 & 51.4 \\
Vanilla RAG   & 56.0 & 100.0 & 10.1 & 55.0 & \cellcolor{ranksecond}{49.2} & 100.0 & 7.5 & 53.8 & \cellcolor{ranksecond}{45.8} & 100.0 & 10.7 & 55.4 \\
\midrule
Internal-Eval & 51.8 & 59.5 & 74.1 & 66.8 & 37.2 & 62.2 & 43.4 & 52.8 & 33.6 & 59.9 & 53.6 & 56.7 \\
Context-Eval  & 52.6 & 54.3 & 88.5 & 71.4 & 44.8 & 86.9 & 30.2 & 58.5 & 41.6 & 83.2 & 42.9 & 63.0 \\
Implicit-SCR  & 58.0 & 93.9 & 28.8 & 61.4 & 38.0 & 62.0 & 22.6 & 42.3 & 32.3 & 58.3 & 23.2 & 40.7 \\
Explicit-SCR  & 44.3 & 47.7 & 79.1 & 63.4 & 22.6 & 30.0 & 67.9 & 49.0 & 17.3 & 23.8 & 53.6 & 38.7 \\
\midrule
TPC           & \cellcolor{ranksecond}{60.9} & 93.9 & 61.9 & \cellcolor{ranksecond}{77.9} & 47.2 & 89.3 & 41.5 & \cellcolor{ranksecond}{65.4} & 43.9 & 89.3 & 58.9 & \cellcolor{ranksecond}{74.1} \\
TACS-LR       & 51.8 & 86.4 & 21.6 & 54.0 & 36.2 & 58.2 & 22.6 & 40.4 & 31.6 & 56.2 & 17.9 & 37.0 \\
\midrule
R-Tuning      & 4.8 & 0.3 & 10.6 & 5.5 & 2.2 & 1.1 & 4.8 & 3.0 & 1.9 & 1.1 & 2.5 & 1.8 \\
CR-DPO        & 46.8 & 43.6 & 89.9 & 66.8 & 29.6 & 40.4 & 69.8 & 55.1 & 27.5 & 41.5 & 71.4 & 56.5 \\
\midrule
\textbf{SABER (Ours)} & \cellcolor{rankfirst}\textbf{62.7} & 91.0 & 83.5 & \cellcolor{rankfirst}\textbf{87.2} & \cellcolor{rankfirst}\textbf{51.3} & 95.8 & 73.6 & \cellcolor{rankfirst}\textbf{84.7} & \cellcolor{rankfirst}\textbf{48.2} & 96.7 & 75.0 & \cellcolor{rankfirst}\textbf{85.9} \\
\bottomrule
\end{tabular}}
\vspace{-2.5mm}
\end{table*}

\begin{table*}[!ht]
\centering
\small
\setlength{\tabcolsep}{2.5pt}
\caption{Per-sub-task breakdown of ConFiQA on the Qwen family, with the same setup as Table~\ref{tab:appendix-main-llama-full}.}
\label{tab:appendix-main-qwen-full}
\resizebox{0.75\textwidth}{!}{%
\begin{tabular}{l|cccc|cccc|cccc}
\toprule
 & \multicolumn{4}{c|}{\cellcolor{gray!15}\textbf{ConFiQA-QA}} & \multicolumn{4}{c|}{\cellcolor{gray!15}\textbf{ConFiQA-MR}} & \multicolumn{4}{c}{\cellcolor{gray!15}\textbf{ConFiQA-MC}} \\
Method & Acc$\uparrow$ & CF & KF & MFS$\uparrow$ & Acc$\uparrow$ & CF & KF & MFS$\uparrow$ & Acc$\uparrow$ & CF & KF & MFS$\uparrow$ \\
\rowcolor{model1}
\midrule
\multicolumn{13}{c}{\textbf{Qwen-2.5-7B}} \\
\midrule
Closed-Book   & 40.2 & 6.0 & 100.0 & 53.0 & 27.2 & 4.2 & 100.0 & 52.1 & 22.8 & 2.3 & 100.0 & 51.2 \\
Vanilla RAG   & 54.8 & 100.0 & 6.3 & 53.1 & 52.8 & 100.0 & 10.9 & 55.4 & 48.8 & 100.0 & 3.3 & 51.6 \\
\midrule
Internal-Eval & 55.8 & 67.0 & 83.6 & 75.3 & 50.4 & 90.5 & 25.0 & 57.8 & 48.5 & 90.5 & 37.4 & 63.9 \\
Context-Eval  & 57.5 & 69.9 & 89.3 & 79.6 & 40.2 & 47.8 & 82.6 & 65.2 & 36.8 & 45.1 & 95.6 & 70.4 \\
Implicit-SCR  & \cellcolor{rankfirst}\textbf{67.0} & 93.5 & 78.6 & \cellcolor{ranksecond}{86.0} & \cellcolor{rankfirst}\textbf{59.0} & 88.7 & 40.2 & 64.4 & \cellcolor{ranksecond}{52.4} & 89.5 & 47.3 & 68.4 \\
Explicit-SCR  & 45.9 & 57.7 & 45.3 & 51.5 & 39.8 & 56.7 & 44.6 & 50.6 & 36.2 & 57.4 & 44.0 & 50.7 \\
\midrule
TPC           & 57.5 & 93.5 & 42.1 & 67.8 & 51.3 & 91.6 & 37.0 & 64.3 & 49.6 & 93.1 & 53.8 & \cellcolor{ranksecond}{73.5} \\
TACS-LR       & 52.8 & 79.2 & 39.6 & 59.4 & 45.2 & 60.2 & 39.1 & 49.6 & 38.2 & 63.3 & 23.1 & 43.2 \\
\midrule
R-Tuning      & 44.4 & 14.0 & 88.2 & 51.1 & 36.3 & 21.2 & 69.2 & 45.2 & 33.3 & 19.2 & 70.0 & 44.6 \\
CR-DPO        & 55.8 & 75.9 & 67.3 & 71.6 & 51.2 & 86.5 & 47.8 & \cellcolor{ranksecond}{67.2} & 47.8 & 86.7 & 51.6 & 69.2 \\
\midrule
\textbf{SABER (Ours)} & \cellcolor{ranksecond}{61.8} & 87.2 & 87.4 & \cellcolor{rankfirst}\textbf{87.3} & \cellcolor{ranksecond}{55.7} & 93.4 & 89.1 & \cellcolor{rankfirst}\textbf{91.3} & \cellcolor{rankfirst}\textbf{53.4} & 94.1 & 95.6 & \cellcolor{rankfirst}\textbf{94.9} \\
\rowcolor{model2}
\midrule
\multicolumn{13}{c}{\textbf{Qwen-2.5-3B}} \\
\midrule
Closed-Book   & 32.5 & 5.4 & 100.0 & 52.7 & 22.8 & 4.0 & 100.0 & 52.0 & 19.2 & 4.7 & 100.0 & 52.3 \\
Vanilla RAG   & 55.5 & 100.0 & 10.8 & 55.4 & 52.7 & 100.0 & 5.2 & 52.6 & 50.0 & 100.0 & 11.5 & 55.8 \\
\midrule
Internal-Eval & 48.3 & 61.1 & 72.5 & 66.8 & 45.9 & 79.2 & 18.8 & 49.0 & 43.2 & 80.1 & 24.4 & 52.2 \\
Context-Eval  & 39.9 & 27.4 & 98.0 & 62.7 & 28.2 & 20.4 & 93.8 & 57.1 & 24.2 & 19.0 & 98.7 & 58.8 \\
Implicit-SCR  & \cellcolor{ranksecond}{57.2} & 91.6 & 42.2 & \cellcolor{ranksecond}{66.9} & \cellcolor{ranksecond}{53.8} & 83.6 & 31.2 & 57.4 & 48.7 & 84.3 & 23.1 & 53.7 \\
Explicit-SCR  & 45.4 & 77.2 & 40.2 & 58.7 & 37.3 & 63.2 & 26.0 & 44.6 & 36.7 & 68.4 & 41.0 & 54.7 \\
\midrule
TPC           & 54.3 & 91.3 & 32.4 & 61.8 & 52.2 & 93.2 & 34.4 & \cellcolor{ranksecond}{63.8} & \cellcolor{ranksecond}{50.6} & 95.1 & 53.8 & \cellcolor{ranksecond}{74.5} \\
TACS-LR       & 50.5 & 90.2 & 15.7 & 53.0 & 45.9 & 78.9 & 20.8 & 49.9 & 41.3 & 76.1 & 15.4 & 45.7 \\
\midrule
R-Tuning      & 39.3 & 17.7 & 79.6 & 48.7 & 45.4 & 38.7 & 80.0 & 59.4 & 40.7 & 34.9 & 66.3 & 50.6 \\
CR-DPO        & 48.6 & 73.4 & 46.1 & 59.7 & 49.0 & 89.7 & 26.0 & 57.9 & 47.4 & 91.3 & 30.8 & 61.1 \\
\midrule
\textbf{SABER (Ours)} & \cellcolor{rankfirst}\textbf{59.1} & 93.5 & 74.5 & \cellcolor{rankfirst}\textbf{84.0} & \cellcolor{rankfirst}\textbf{56.1} & 94.8 & 69.8 & \cellcolor{rankfirst}\textbf{82.3} & \cellcolor{rankfirst}\textbf{52.7} & 93.7 & 87.2 & \cellcolor{rankfirst}\textbf{90.4} \\
\bottomrule
\end{tabular}}
\vspace{-2mm}
\end{table*}

\subsection{Threshold sweep on the abstention experiment}
\label{sec:appendix-tau-sweep}

Table~\ref{tab:appendix-tau-sweep} reports the full threshold sweep used to produce Figure~\ref{fig:abstain-rc-curve}. For each backbone, we sweep $\tau$ over $[0.05, 0.95]$ in steps of $0.05$ and report Score, Coverage (Cov), Risk on answered ($R_C$), and abstention F1 ($F_1$), all in percent.

\begin{table}[!ht]
\centering
\scriptsize
\caption{Full $\tau$-sweep for SABER's selective-answering decision rule across four backbone.}
\label{tab:appendix-tau-sweep}
\setlength{\tabcolsep}{3pt}
\begin{tabular}{c|rrrr|rrrr|rrrr|rrrr}
\toprule
 & \multicolumn{4}{c|}{\textbf{Llama-3.1-8B}} & \multicolumn{4}{c|}{\textbf{Qwen-2.5-7B}} & \multicolumn{4}{c|}{\textbf{Llama-3.2-3B}} & \multicolumn{4}{c}{\textbf{Qwen-2.5-3B}} \\
$\tau$ & Score & Cov & $R_C$ & $F_1$ & Score & Cov & $R_C$ & $F_1$ & Score & Cov & $R_C$ & $F_1$ & Score & Cov & $R_C$ & $F_1$ \\
\midrule
0.05 & 38.4 & 84.0 & 27.1 & 53.4 & 32.2 & 85.1 & 31.1 & 50.7 & 29.0 & 81.6 & 32.2 & 54.8 & 29.4 & 83.0 & 32.3 & 51.3 \\
0.10 & 40.2 & 81.2 & 25.2 & 58.8 & 34.8 & 80.9 & 28.5 & 58.2 & 32.4 & 76.6 & 28.8 & 62.8 & 32.2 & 79.3 & 29.7 & 58.1 \\
0.15 & 41.1 & 79.2 & 24.0 & 61.5 & 37.0 & 77.6 & 26.2 & 63.8 & 34.4 & 73.5 & 26.6 & 67.0 & 34.0 & 76.1 & 27.7 & 62.7 \\
0.20 & 41.7 & 77.8 & 23.2 & 63.3 & 39.0 & 74.3 & 23.7 & 68.5 & 35.5 & 71.0 & 25.0 & 69.8 & 35.6 & 73.5 & 25.8 & 66.2 \\
0.25 & 42.4 & 76.0 & 22.1 & 65.1 & 40.3 & 71.9 & 22.0 & 71.2 & 36.4 & 68.5 & 23.4 & 71.6 & 36.9 & 71.2 & 24.1 & 69.2 \\
0.30 & 43.1 & 74.3 & 21.0 & 66.8 & 41.7 & 69.6 & 20.0 & 73.4 & 37.2 & 66.2 & 21.9 & 73.3 & 37.8 & 69.0 & 22.6 & 71.2 \\
0.35 & 43.6 & 72.5 & 19.9 & 68.6 & 42.5 & 67.4 & 18.5 & 75.0 & 37.7 & 64.4 & 20.8 & 74.6 & 38.7 & 67.0 & 21.1 & 73.1 \\
0.40 & 44.0 & 70.5 & 18.8 & 70.3 & 43.0 & 64.9 & 16.9 & 76.3 & 38.0 & 62.5 & 19.6 & 75.6 & 39.2 & 65.2 & 19.9 & 74.4 \\
0.45 & 44.0 & 68.3 & 17.8 & 71.2 & 43.4 & 62.3 & 15.2 & 77.3 & 38.6 & 60.5 & 18.1 & 76.4 & 39.7 & 62.6 & 18.3 & 75.6 \\
\textbf{0.50} & \textbf{43.9} & \textbf{65.1} & \textbf{16.3} & \textbf{72.5} & \textbf{43.4} & \textbf{58.5} & \textbf{13.0} & \textbf{78.0} & \textbf{38.5} & \textbf{58.6} & \textbf{17.2} & \textbf{76.8} & \textbf{39.6} & \textbf{60.0} & \textbf{17.0} & \textbf{76.6} \\
0.55 & 43.7 & 62.6 & 15.1 & 72.9 & 42.7 & 56.2 & 12.0 & 77.8 & 38.0 & 56.0 & 16.1 & 76.9 & 39.5 & 56.2 & 14.9 & 77.1 \\
0.60 & 43.3 & 60.8 & 14.3 & 73.1 & 42.4 & 54.3 & 10.9 & 78.0 & 38.0 & 53.9 & 14.8 & 77.4 & 39.3 & 53.9 & 13.6 & 77.8 \\
0.65 & 42.9 & 58.9 & 13.6 & 73.2 & 41.9 & 52.2 & 9.9 & 78.0 & 37.4 & 51.9 & 14.0 & 77.5 & 38.8 & 51.6 & 12.4 & 78.1 \\
0.70 & 42.5 & 57.0 & 12.7 & 73.6 & 41.2 & 50.3 & 9.0 & 77.7 & 37.2 & 50.0 & 12.8 & 77.8 & 38.5 & 49.4 & 11.0 & 78.2 \\
0.75 & 41.9 & 55.2 & 12.0 & 73.5 & 40.0 & 48.2 & 8.4 & 77.2 & 36.4 & 47.9 & 11.9 & 77.7 & 37.6 & 47.3 & 10.2 & 78.1 \\
0.80 & 41.4 & 53.0 & 11.0 & 73.3 & 39.3 & 45.9 & 7.1 & 76.8 & 35.7 & 45.4 & 10.6 & 77.7 & 36.4 & 44.4 & 8.9 & 77.9 \\
0.85 & 40.2 & 50.2 & 10.0 & 73.2 & 38.0 & 43.3 & 6.1 & 76.2 & 34.6 & 42.9 & 9.7 & 77.3 & 34.8 & 41.2 & 7.7 & 77.6 \\
0.90 & 38.3 & 46.9 & 9.1 & 72.5 & 35.7 & 40.1 & 5.5 & 74.9 & 33.2 & 40.0 & 8.5 & 76.8 & 33.4 & 37.9 & 6.0 & 77.2 \\
0.95 & 34.8 & 41.2 & 7.8 & 71.0 & 32.3 & 35.5 & 4.5 & 73.0 & 30.4 & 35.8 & 7.5 & 75.4 & 29.8 & 32.4 & 4.1 & 75.6 \\
\bottomrule
\end{tabular}
\vspace{-3mm}
\end{table}

\subsection{Layer-grid heatmap on the Qwen family}
\label{sec:appendix-qwen-heatmap}

\begin{figure}[!ht]
\centering
\includegraphics[width=\linewidth]{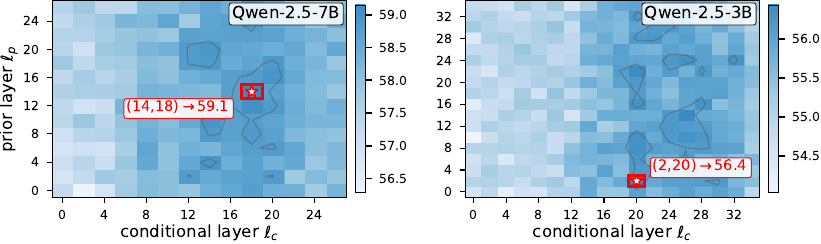}
\caption{Layer sensitivity of SABER on the Qwen family (counterpart of Figure~\ref{fig:grid2d-acc-llama}). Each panel shows end-to-end answer accuracy (\%) over the $(\ell_p, \ell_c)$ grid at stride~$2$. Darker cells denote higher accuracy and the grey contour traces the 90th-percentile region. The red square marks the selected layer pair.}
\label{fig:appendix-grid2d-qwen}
\vspace{-3mm}
\end{figure}

Figure~\ref{fig:appendix-grid2d-qwen} shows the Qwen counterpart of Figure~\ref{fig:grid2d-acc-llama} on the two Qwen backbones. Both panels exhibit a broad high-accuracy plateau around the chosen layer pair, with the optimum on Qwen-2.5-7B falling in the mid-to-late depth band as on the Llama family, while Qwen-2.5-3B selects an unusually shallow prior layer ($\ell_p{=}2$) that nevertheless lies inside its own plateau region.

\subsection{Cross-Dataset (Leave-One-Dataset-Out) Generalization}
\label{sec:appendix-lodo}

The main-results setup trains SABER on the union of the seven datasets' training partitions and evaluates on each dataset's test split, so every dataset contributes both train and test instances. This subsection asks a complementary question of how well SABER transfers to a dataset whose distribution is unseen during training. We adopt a leave-one-dataset-out (LODO) protocol. For each held-out dataset, SABER is retrained on the train, validation, and test instances of the other six datasets, then evaluated on the held-out test split. Each fold uses roughly 53K--67K training instances, comparable to the main-results training pool, but disjoint from the held-out test split in dataset identity. We report results for every backbone--held-out dataset cell.

\textbf{Transfer pattern.}
Table~\ref{tab:appendix-lodo} reports per-cell IID versus LODO Acc and MFS, showing that SABER's belief estimator transfers robustly across datasets. On the five typical datasets (ConFiQA-MR, ConFiQA-MC, ConFiQA-QA, TriviaQA, and NQ), the LODO Acc gap stays within roughly \(\pm 3\%\) and the MFS gap within roughly \(10\%\) across all four backbones, indicating that the learned PK/CK arbitration behaviour is largely dataset-agnostic rather than relying on dataset-specific surface forms. The main exceptions are ConflictBank and ConflictQA-PopQA, where the larger drops are explained by training-pool composition. These two datasets carry distinctive conflict formats, namely long synthetic claim-evidence templates in ConflictBank and engineered counterfactual injections in ConflictQA-PopQA, which are absent from the remaining six-dataset pool used to train each LODO fold. As a result, LODO SABER has no near-distribution conflict signal to draw on for these specific folds, and the gap is expected to close once additional conflict-style datasets enter the pool.

\begin{table}[!ht]
\centering
\small
\caption{Cross-dataset (LODO) generalization of SABER. For each held-out dataset, SABER is trained on the union of the other six datasets' train, validation, and test instances, and evaluated on the held-out test split. IID denotes SABER trained on the training partitions of all seven datasets. $\Delta$ reports LODO minus IID in percentage points.}
\label{tab:appendix-lodo}
\setlength{\tabcolsep}{4pt}
\begin{tabular}{ll|cc|cc|cc}
\toprule
& & \multicolumn{2}{c|}{\textbf{IID}} & \multicolumn{2}{c|}{\textbf{LODO}} & \multicolumn{2}{c}{\textbf{$\Delta$ (\%)}} \\
\textbf{Backbone} & \textbf{Held-out} & Acc & MFS & Acc & MFS & $\Delta$Acc & $\Delta$MFS \\
\midrule
Llama-3.1-8B & ConFiQA-QA & 68.5 & 79.6 & 65.5 & 72.9 & -3.0 & -6.6 \\
 & ConFiQA-MR & 58.2 & 74.7 & 56.9 & 74.3 & -1.2 & -0.5 \\
 & ConFiQA-MC & 51.1 & 87.2 & 50.7 & 82.7 & -0.3 & -4.5 \\
 & ConflictQA & 53.6 & 90.5 & 46.5 & 65.6 & -7.1 & -24.9 \\
 & ConflictBank & 79.7 & 88.5 & 54.3 & 65.3 & -25.4 & -23.2 \\
 & TriviaQA & 83.9 & 76.4 & 82.2 & 85.1 & -1.7 & +8.7 \\
 & NQ & 54.5 & 72.8 & 51.2 & 63.3 & -3.3 & -9.5 \\
\midrule
Qwen-2.5-7B & ConFiQA-QA & 63.1 & 89.8 & 55.8 & 71.5 & -7.3 & -18.3 \\
 & ConFiQA-MR & 56.2 & 91.0 & 55.2 & 81.9 & -1.1 & -9.1 \\
 & ConFiQA-MC & 53.7 & 92.6 & 52.8 & 88.9 & -0.9 & -3.7 \\
 & ConflictQA & 49.4 & 93.0 & 46.0 & 68.3 & -3.3 & -24.7 \\
 & ConflictBank & 74.7 & 86.3 & 52.3 & 64.5 & -22.4 & -21.8 \\
 & TriviaQA & 71.7 & 89.5 & 67.1 & 80.7 & -4.6 & -8.8 \\
 & NQ & 53.3 & 79.1 & 51.7 & 78.1 & -1.7 & -1.0 \\
\midrule
Llama-3.2-3B & ConFiQA-QA & 64.0 & 90.6 & 56.1 & 72.9 & -7.9 & -17.7 \\
 & ConFiQA-MR & 51.3 & 85.5 & 51.6 & 78.7 & +0.3 & -6.8 \\
 & ConFiQA-MC & 48.0 & 88.2 & 47.4 & 79.0 & -0.6 & -9.2 \\
 & ConflictQA & 47.1 & 91.5 & 43.0 & 67.9 & -4.2 & -23.6 \\
 & ConflictBank & 70.5 & 87.3 & 59.6 & 66.9 & -10.8 & -20.4 \\
 & TriviaQA & 76.2 & 70.6 & 69.2 & 72.7 & -7.1 & +2.1 \\
 & NQ & 50.8 & 71.0 & 47.9 & 64.5 & -2.9 & -6.5 \\
\midrule
Qwen-2.5-3B & ConFiQA-QA & 59.5 & 85.7 & 49.8 & 65.3 & -9.7 & -20.4 \\
 & ConFiQA-MR & 55.7 & 80.0 & 55.8 & 76.6 & +0.1 & -3.4 \\
 & ConFiQA-MC & 52.5 & 88.4 & 52.8 & 87.7 & +0.3 & -0.8 \\
 & ConflictQA & 44.4 & 92.1 & 41.0 & 73.1 & -3.4 & -19.1 \\
 & ConflictBank & 72.3 & 90.0 & 64.3 & 56.2 & -8.1 & -33.9 \\
 & TriviaQA & 66.2 & 88.8 & 60.8 & 78.7 & -5.4 & -10.1 \\
 & NQ & 50.4 & 73.5 & 48.8 & 74.6 & -1.7 & +1.0 \\
\bottomrule
\end{tabular}
\end{table}

\textbf{Comparison against the strongest zero-shot baseline.}
We further ask whether LODO SABER can outperform the strongest prompt-based zero-shot baseline on each held-out cell. To make the comparison maximally challenging, we select, for every (backbone, held-out dataset) cell, the best single prompt-based baseline and compare against its paired (Acc, MFS). Each row therefore uses one real deployable baseline rather than a per-metric oracle, and the baseline enters the comparison at the operating point most favorable to conflict faithfulness.

Table~\ref{tab:appendix-lodo-vs-baselines} shows that SABER simultaneously dominates on both axes even without seeing the held-out distribution during training. It outperforms the strongest baseline on Acc in \textbf{19/28} cells and on MFS in \textbf{20/28} cells, with MFS gains frequently in the \(15\%\)--\(30\%\) range. Several cells show large simultaneous gains on both metrics, including \(+28.7\) Acc and \(+28.8\) MFS on Qwen-2.5-3B / ConFiQA-MC, \(+19.6\) Acc and \(+20.7\) MFS on Llama-3.1-8B / ConFiQA-MC, and \(+18.9\) Acc and \(+15.2\) MFS on Llama-3.1-8B / TriviaQA.

This comparison exposes a fundamental trade-off in prompt-only methods. Baselines that achieve high MFS, typically the PK-leaning Context-Eval and Internal-Eval, do so by sacrificing Acc, while Acc-oriented baselines collapse on MFS. SABER avoids this trade-off by learning path-level reliability beliefs and remains strong on both Acc and MFS even when trained on datasets disjoint from the held-out fold. The remaining losses concentrate on stylistically idiosyncratic folds, most notably the larger ConFiQA-QA cells on the Qwen family and the ConflictBank fold, where the six-dataset LODO training pool lacks near-distribution signals for those specific surface patterns.

\begin{table}[!ht]
\centering
\small
\caption{LODO SABER vs. the strongest prompt-based zero-shot baseline on each backbone--held-out cell. For each cell, we select the best prompt-based baseline among Closed-Book (CB), Vanilla RAG (RAG), Internal-Eval (Int-Eval), Context-Eval (Ctx-Eval), Implicit-SCR (Imp-SCR), and Explicit-SCR (Exp-SCR), and report its paired Acc and MFS. Bold marks SABER wins, and \(\Delta\) is reported in percent. LODO SABER beats the best baseline on 19/28 cells in Acc and 20/28 cells in MFS, showing that it remains strong on both metrics while prompt-only baselines often trade one for the other.}
\label{tab:appendix-lodo-vs-baselines}
\setlength{\tabcolsep}{4pt}
\begin{tabular}{ll|l|cc|cc|cc}
\toprule
& & & \multicolumn{2}{c|}{\textbf{Top-MFS Baseline}} & \multicolumn{2}{c|}{\textbf{LODO SABER}} & \multicolumn{2}{c}{\textbf{$\Delta$ (\%)}} \\
\textbf{Backbone} & \textbf{Held-out} & \textbf{Method} & Acc & MFS & Acc & MFS & $\Delta$Acc & $\Delta$MFS \\
\midrule
Llama-3.1-8B & ConFiQA-QA & Imp-SCR & 66.8 & 73.7 & 65.5 & 72.9 & -1.3 & -0.8 \\
 & ConFiQA-MR & RAG & 57.3 & 59.3 & 56.9 & 74.3 & -0.4 & \textbf{+15.0} \\
 & ConFiQA-MC & Exp-SCR & 31.2 & 62.0 & 50.7 & 82.7 & \textbf{+19.6} & \textbf{+20.7} \\
 & ConflictQA & Exp-SCR & 40.9 & 63.1 & 46.5 & 65.6 & \textbf{+5.6} & \textbf{+2.5} \\
 & ConflictBank & Ctx-Eval & 58.9 & 76.0 & 54.3 & 65.3 & -4.6 & -10.7 \\
 & TriviaQA & Exp-SCR & 63.3 & 69.8 & 82.2 & 85.1 & \textbf{+18.9} & \textbf{+15.2} \\
 & NQ & Exp-SCR & 46.3 & 63.0 & 51.2 & 63.3 & \textbf{+5.0} & \textbf{+0.2} \\
\midrule
Qwen-2.5-7B & ConFiQA-QA & Imp-SCR & 67.0 & 86.0 & 55.8 & 71.5 & -11.2 & -14.5 \\
 & ConFiQA-MR & Ctx-Eval & 40.2 & 65.2 & 55.2 & 81.9 & \textbf{+15.0} & \textbf{+16.7} \\
 & ConFiQA-MC & Ctx-Eval & 36.8 & 70.4 & 52.8 & 88.9 & \textbf{+15.9} & \textbf{+18.5} \\
 & ConflictQA & Int-Eval & 38.2 & 67.6 & 46.0 & 68.3 & \textbf{+7.8} & \textbf{+0.7} \\
 & ConflictBank & Ctx-Eval & 60.7 & 62.6 & 52.3 & 64.5 & -8.4 & \textbf{+1.9} \\
 & TriviaQA & Int-Eval & 65.4 & 78.0 & 67.1 & 80.7 & \textbf{+1.7} & \textbf{+2.7} \\
 & NQ & Imp-SCR & 52.5 & 64.5 & 51.7 & 78.1 & -0.8 & \textbf{+13.6} \\
\midrule
Llama-3.2-3B & ConFiQA-QA & Ctx-Eval & 52.6 & 71.4 & 56.1 & 72.9 & \textbf{+3.5} & \textbf{+1.5} \\
 & ConFiQA-MR & Ctx-Eval & 44.8 & 58.5 & 51.6 & 78.7 & \textbf{+6.8} & \textbf{+20.2} \\
 & ConFiQA-MC & Ctx-Eval & 41.6 & 63.0 & 47.4 & 79.0 & \textbf{+5.8} & \textbf{+15.9} \\
 & ConflictQA & Int-Eval & 37.2 & 70.7 & 43.0 & 67.9 & \textbf{+5.8} & -2.8 \\
 & ConflictBank & Ctx-Eval & 49.9 & 71.9 & 59.6 & 66.9 & \textbf{+9.7} & -5.0 \\
 & TriviaQA & Int-Eval & 72.1 & 69.9 & 69.2 & 72.7 & -2.9 & \textbf{+2.8} \\
 & NQ & Int-Eval & 43.8 & 61.8 & 47.9 & 64.5 & \textbf{+4.1} & \textbf{+2.8} \\
\midrule
Qwen-2.5-3B & ConFiQA-QA & Imp-SCR & 57.2 & 66.9 & 49.8 & 65.3 & -7.5 & -1.6 \\
 & ConFiQA-MR & Imp-SCR & 53.8 & 57.4 & 55.8 & 76.6 & \textbf{+2.0} & \textbf{+19.2} \\
 & ConFiQA-MC & Ctx-Eval & 24.2 & 58.8 & 52.8 & 87.7 & \textbf{+28.7} & \textbf{+28.8} \\
 & ConflictQA & Int-Eval & 30.6 & 61.4 & 41.0 & 73.1 & \textbf{+10.4} & \textbf{+11.7} \\
 & ConflictBank & Ctx-Eval & 39.6 & 59.0 & 64.3 & 56.2 & \textbf{+24.7} & -2.8 \\
 & TriviaQA & Int-Eval & 63.3 & 82.6 & 60.8 & 78.7 & -2.5 & -3.8 \\
 & NQ & Int-Eval & 41.7 & 71.0 & 48.8 & 74.6 & \textbf{+7.0} & \textbf{+3.6} \\
\bottomrule
\end{tabular}
\end{table}

\section{Computational Resources}
\label{sec:appendix-compute}

SABER's predictor trains roughly an order of magnitude faster than R-Tuning and two orders faster than CR-DPO, while running on smaller hardware. All experiments were conducted on NVIDIA A5000 GPUs with 24~GB memory. Table~\ref{tab:appendix-train-cost} reports per-backbone training cost for the three trained methods. Compared to the two fine-tuning baselines, SABER offers four practical advantages:

\begin{itemize}
\item \textbf{Smaller predictor, smaller memory footprint.} SABER's MLP predictor uses $\sim$5M trainable parameters and fits within $<$2~GB peak GPU memory, against $\sim$14M LoRA parameters and 5--10~GB peak memory for R-Tuning and CR-DPO.

\item \textbf{Single-GPU training.} SABER trains on a single A5000, whereas CR-DPO and R-Tuning require multi-GPU DDP (4--8 GPUs). This makes SABER accessible to typical academic compute budgets without specialised infrastructure.

\item \textbf{An order of magnitude cheaper than R-Tuning, two orders cheaper than CR-DPO.} Across the four backbones, SABER's total predictor training cost is $<$0.6 GPU-hours, against $\sim$13 GPU-hours for R-Tuning ($\sim$22$\times$) and $\sim$72 GPU-hours for CR-DPO ($\sim$120$\times$). Predictor training and evaluation each take less than 0.2 GPU-hour, even for 8B backbones.

\item \textbf{Amortised one-time setup, no checkpoint maintenance.} SABER's main up-front cost is hidden-state extraction across the four backbones (approximately 15 GPU-hours of effective compute for the production run, since multiple extraction jobs share each GPU and inference rarely saturates the device), paid once per backbone and amortised across every probe variant; the backbone itself stays frozen and reusable. R-Tuning and CR-DPO instead pay their full training cost whenever the backbone, dataset, or hyperparameters change, and must maintain a forked checkpoint alongside the base model.
\end{itemize}

Taken together, even when the one-time hidden-state extraction is included in the budget, SABER's total effective compute (approximately 15 GPU-hours) is substantially less than CR-DPO ($\sim$72 GPU-hours, $\sim$5$\times$ less) and comparable to R-Tuning ($\sim$13 GPU-hours), while running on a single A5000 at a fraction of either baseline's memory footprint. Once the cached states are in place, every subsequent probe re-training takes minutes rather than hours, so the total compute gap widens further as soon as more than one probe configuration is explored. We therefore view SABER as the most practical of the three trained approaches to deploy and iterate on, particularly for groups without large-scale multi-GPU infrastructure.

\begin{table}[!ht]
\centering
\small
\caption{Per-backbone training cost on NVIDIA A5000 (24~GB). SABER's one-time hidden-state extraction (not shown) costs $\sim$5 GPU-h per 8B backbone and $\sim$3 GPU-h per 3B backbone ($\sim$15 GPU-h total of effective compute for the production run, with multiple extraction jobs sharing each GPU), amortised across every variant trained on the cached states.}
\label{tab:appendix-train-cost}
\begin{tabular}{l|ccc}
\toprule
& \textbf{CR-DPO} & \textbf{R-Tuning} & \textbf{SABER predictor} \\
\midrule
Trainable params & $\sim$14M (LoRA) & $\sim$14M (LoRA) & $\sim$5M (MLP) \\
Peak GPU memory  & $\sim$10 GB & $\sim$5 GB & $<$2 GB \\
Hardware (DDP)   & 8$\times$A5000 & 4--8$\times$A5000 & 1$\times$A5000 \\
\midrule
GPU-h on Llama-3.1-8B & 21 & 5.3 & $<$0.2 \\
GPU-h on Llama-3.2-3B & 14 & 2.3 & $<$0.1 \\
GPU-h on Qwen-2.5-7B  & 23 & 2.3 & $<$0.2 \\
GPU-h on Qwen-2.5-3B  & 14 & 3.0 & $<$0.1 \\
\midrule
Total GPU-h & $\sim$72 & $\sim$13 & $<$0.6 \\
\bottomrule
\end{tabular}
\end{table}

\section{Limitations}
\label{sec:appendix-limitations}

We discuss several limitations of our study, primarily as scope choices that motivate future work.

\paragraph{Backbone scope.} Our experiments cover four open-weight instruction-tuned backbones from the Llama family, Llama-3.1-8B and Llama-3.2-3B, and the Qwen family, Qwen-2.5-7B and Qwen-2.5-3B. Extending SABER to substantially larger backbones, where hidden states may exhibit different self-awareness profiles, is left to future work.

\paragraph{Retrieval scope.} Our study assumes that the retrieved context has already been supplied and focuses on the well-defined problem of resolving knowledge conflict between this context and the model's parametric knowledge. We do not address the upstream retrieval process itself, including how candidate passages are selected, ranked, or pre-filtered from a corpus, which is an orthogonal line of work.

\paragraph{Static benchmark.} Our benchmark labels four fixed LLMs at a single point in time. As new LLMs are released, per-backbone PK- and CK-path correctness labels need to be re-collected. The labeling pipeline in Eq.~\ref{eq:answer-paths}--\ref{eq:correctness-labels} is fully reproducible, but incurs non-trivial inference cost for each new backbone.

\section{Broader Impacts}
\label{sec:appendix-broader-impacts}

SABER aims to make retrieval-augmented generation more honest by enabling calibrated abstention when neither the parametric nor the contextual answer path is reliable. Its intended positive impact is to reduce confidently wrong outputs in LLM-based question answering. We do not anticipate substantial negative societal impacts specific to our method.

\clearpage

\end{document}